\documentclass[aps,pre,amsmath,amssymb,twocolumn,showpacs,byrevtex,superscriptaddress]{revtex4}
\usepackage{graphicx}
\begin{document}

\author{Emanuela Zaccarelli}
\address{Dipartimento di Fisica and CNR-INFM-SOFT, Universit\`a di Roma La Sapienza, Piazzale A. Moro 2, I-00185, Rome, Italy}
\author{Francesco Sciortino}
\address{Dipartimento di Fisica and CNR-INFM-SOFT, Universit\`a di Roma La Sapienza, Piazzale A. Moro 2, I-00185, Rome, Italy}
\author{Piero Tartaglia}
\address{Dipartimento di Fisica and CNR-INFM-SMC, Universit\`a di Roma La Sapienza, Piazzale A. Moro 2, I-00185, Rome, Italy}
\title{A spherical model with directional interactions: \\ I. Static properties}

\date{\today}
% This is useful both in the context of patchy colloids and of network-forming liquids.  The recipe provided here can be straightforward generalise to a desired geometrical arrangement and coordination number.
\begin{abstract}
We introduce a simple spherical model whose structural properties are
similar to the ones generated by models with directional interactions,
by employing a binary mixture of large and small hard spheres, with a
square-well attraction acting only between particles of different
size. The small particles provide the bonds between the large ones.
With a proper choice of the interaction parameters, as well as of the
relative concentration of the two species, it is possible to control
the effective valence.  Here we focus on a specific choice of the
parameters which favors tetrahedral ordering and study the equilibrium
static properties of the system in a large window of densities and
temperatures. Upon lowering the temperature we observe a progressive
increase in local order, accompanied by the formation of a
four-coordinated network of bonds.  Three different density regions
are observed: at low density the system phase separates into a gas and
a liquid phase; at intermediate densities a network of fully bonded
particles develops; at high densities --- due to the competition
between excluded volume and attractive interactions --- the system
forms a defective network.  The very same behavior has been previously
observed in numerical studies of non-spherical models for molecular
liquids, such as water, and in models of patchy colloidal particles.
Differently from these models, theoretical treatments devised for
spherical potentials, e.g. integral equations and ideal mode coupling
theory for the glass transition can be applied in the present case,
opening the way for a deeper understanding of the thermodynamic and
dynamic behavior of low valence molecules and particles.  
%% describe patchy colloidal particles with
%% attractive directional interactions.  The model is relevant to the
%% study of gelation in attractive colloids, without an intervening phase
%% separation, and to the connection with network-forming liquids.  We
%% focus on the case where particles form aggregate in a tetrahedral
%% geometry.  Increasing the attraction strength, the aggregates form a
%% tetrahedral network, as it is observed in the static structure
%% factor. We discuss the connections of the model with existing
%% limeted-valency or primitive models and with experimental systems.

%% Contrarily to maximum-valency type models 
%% studied previously, many-body interactions are not present in 
%% the effective interaction potential,
%% so that integral equation theories can be also applied to the system.
%% By numerical simulations, we confirm dynamical results as for the existing
%% models. Moreover, we observe a characteristic
%% structure factor behaviour typical of tetrahedral molecules, such as silica.
%% Finally we discuss connections with experimental systems.
\end{abstract}

% 82.70.Gg Gels and sols
% 61.20.Lc Structure of liquids; time-dependent properties; relaxation
% 61.43.Hv fractals, macroscopic aggregates including DLA
% 82.70.Dd Colloids
% 64.60.Ak Renorm. group, fractal, and percolation studies of phase tranistions
% 64.70.Pf Glass transitions

\maketitle

\section{Introduction}

The study of dynamic arrest in atomic and molecular systems is an
active field of research\cite{kobbook}.  Close to the glass
transition, in a small window of values of the external parameters
(temperature or density/pressure) the dynamics of the system slows
down by fifteen or more orders of magnitude. The slowing down of the
dynamics varies in the temperature or density/pressure dependence from
an Arrhenius law for the so-called strong network glass-forming
liquids to a more complex super-Arrhenius behavior for fragile
glass-forming liquids\cite{Ang85a}.  Signatures of the
slowing down of the dynamics are observed
at temperatures higher (or density lower) than the calorimetric glass
transition and have been interpreted as the genuine precursor of the
arrest phenomenon by the ideal Mode Coupling Theory
(MCT)\cite{goetze}.  The study of the dynamics in colloidal
systems\cite{advances,Zac07a} has added new fuel to the discussion on
dynamic arrest.  Experimental realizations of simple models amenable
to theoretical investigation has considerably boosted the
understanding of the glass phenomenon and opened up an ampler view by
adding to the arena the gelation issue, i.e.  the possibility of
observing arrest at low density, driven by the formation of
inter-particle attractive bonds.  Indeed, nowadays it is possible to
realize colloidal systems which closely follow the hard-sphere (HS)
equation of state, as well as to control an additional attractive
interaction, both in range and in strength, moving the field of
colloidal liquids towards molecular systems\cite{Voigt06}.

Recent studies have shown that when the hard-core is complemented by a
spherical attractive potential, phase separation preempts the
possibility of continuously approaching the slowing down of the
dynamics at low density.  For Lennard-Jones particles,
Sastry\cite{Sas00PRL} showed that the glass line intersects the
liquid-gas spinodal on the liquid branch, suggesting that homogeneous
arrested states are only possible for significantly large densities.
More recently, the same scenario has been shown to hold even in the
limit of very short-ranged spherical
attractions\cite{Zaccapri,Fof05a}, down to the Baxter limit
(infinitesimal attraction range).  On the other hand, a series of
studies have suggested that a progressive arrest at small packing
fraction can be observed if the inter-particle interactions are highly
directional (patchy interactions), when the effective valence becomes
small\cite{Zaccagel}. On decreasing the valence below six, the
gas-liquid unstable region progressively shrinks to smaller densities,
opening up an intermediate region where a stable network of bonded
particles forms. The shrinking of the unstable region can be tuned
continuously down to a vanishing densities on decreasing the valence
down to two\cite{Bianchi_06}. When the average valence is slightly
larger than two, it is possible to observe {\it empty liquids}, i.e.
states with a temperature smaller than the critical temperature but
with an extremely small liquid-state density.  Numerical studies of
the dynamics of liquids with patchy interactions have shown that the
bond lifetime of the liquid network progressively increases upon
cooling, providing evidence of the possibility of approaching arrested
states continuously. The slowing down of the dynamics in the newly
accessible density region follows an Arrhenius
law\cite{Zac06JCP,DeMichele_05,DeMichele_06}.  The arrest transition
can be simultaneously interpreted in terms of {\it equilibrium
gelation} in the colloid community\cite{Zac06JCP,Zac07a} and of {\it
strong-glass} formation in the supercooled liquids
community\cite{DeMichele_05,DeMichele_06}.  These studies have clearly
shown that reduction of valence is an essential ingredient for
extending the glass line from the high-density/pressure region,
dominated by repulsion, to intermediate regions where attraction and
repulsion cooperate, down to the region of lower densities where
attraction become the dominant arrest mechanism.

While simulations are a well documented method for studying the static
and dynamic properties of non-spherical models, theoretical approaches
are still not well developed, especially when the geometry of the
bonds is such that inter-particle correlations propagate over several
neighbors.  The situation is even worse for microscopic theories of
the glass transition, requiring structural quantities as input
data. For the case of molecules,
molecular\cite{Schi97a,Fab99b,Schi00b,Schi00a} and
site-site\cite{Cho98a,chonggotze,Cho04a} extensions of MCT
have been developed, but their use has been limited due to the nature
of the approximations and to the complexity of the calculations. An
effective spherical approximation for non-spherical potentials has
also been recently reported\cite{yatsenko}.  For
these reasons, the theoretical evaluation of the MCT glass line for
patchy non-spherical models has never been attempted so far.  For the
case of water (at ambient pressure) a molecular-MCT calculation has
been reported\cite{Fab99b,Fab00a}.  Besides water, the only network
forming system whose slow dynamic properties have been investigated
--- starting from appropriate structural properties --- is
silica\cite{Sci01aPRL,nauroth,Verrocchio}. In the
investigated model of silica, the network is formed by a binary
mixture of two spherically interacting particles and hence all the
complications associated to angular constraints are missing.
Theoretical studies of silica have been limited to one specific value
of the density and no clear picture of the glass-line in the
$T$-$\rho$ plane has been provided.

As a first step toward a simple network-forming model for which
theoretical calculations of the location of the arrest line in the
phase diagram are in principle feasible, we introduce here a simple
binary mixture model of particles interacting via short-range square
well potentials and which is able to reproduce the same pattern (both
for structural and dynamic quantities) of the previously studied
non-spherical patchy models. The idea is borrowed from existing models
for silica based on pair-wise additive interactions\cite{bks} but with
a much simpler interaction potential. Due to the absence of the
long-range electrostatic interactions, the model can be simulated in
the entire phase diagram and the slowing down of the dynamics can be
followed over a window of more than five orders of magnitude.  The
present article provides an evaluation of the structural properties of
this model, based on extensive event-driven molecular dynamics
simulations.  A future companion article\cite{prossimo} will report
the dynamic properties of the model. The manuscript is organized as
follows: section II introduces the model; section III reports results
for the structural, energetic and geometric properties . Section IV
discusses the phase diagram, complemented by the study of various
thermodynamic loci, while section V is devoted to conclusions.

\section{The model}
Previously studied simple models for network forming liquids are based
on many-body interactions\cite{Zaccagel,LargoPRE} or angular
constraints\cite{Kern_03,Bianchi_06}.  Here we introduce a simple
model retaining both pair-wise additivity and spherical interactions,
but which is capable of producing geometrical arrangement into a
locally ordered network structure.  The oxymoron spherical model with
directional interactions is realised through the introduction of a
second species in the mixture that represents a sort of floating
bond. In this way, an effective one-component directional potential
for the colloidal particles is obtained.

In more details, we consider $N_1$ hard sphere colloids (for which we
reserve the name {\it particles} in the following) with diameter
$\sigma_{11}$.  We mix them with $N_2$ small particles with
hard-sphere diameter $\sigma_{22}$ (named {\it floating bonds} in the
following). Particles and floating bonds interact  via a non-additive
(NA), short-ranged, attractive square-well (SW) potential of depth
$-u_0$ and range $\delta$.  Thus, the floating bonds link particles
providing a connection between them.  More precisely the interaction
potential is
\begin{eqnarray}
V_{11}(r)&=&\left\{ \begin{array}{c} \infty \ \ r<\sigma_{11}\\ 
0 \ \ r>\sigma_{11}\end{array}
\nonumber \right. \\
V_{12}(r)&=&\left\{ \begin{array}{c} \!\!\!\!\!\!\!\!\!\!\!\!\!\!\!\!\!\!\!
\infty \ \ \ \ \ \ \ \ r<\sigma_{12}^{NA} \\ 
-u_0 \ \ \sigma_{12}^{NA}<r< \sigma_{12}^{NA}+\delta 
\\ \!\!\!\!\!\!\!\!\!\!0 \ \ \ \ \ \ \ \ r> \sigma_{12}^{NA}+\delta \end{array} 
\right.
\nonumber \\
V_{22}(r)&=&\left\{ \begin{array}{c} \infty \ \ r<\sigma_{22} \\ 0 \ \ r>
\sigma_{22} \end{array} \right.
\end{eqnarray}

We choose the potential parameters in such a way that (i) each
floating bond binds no more than two particles; and (ii) the maximum
number of floating bonds binding to a single particle is fixed,
providing the valence of the model.  To enforce condition (i), we
recall that three identical touching hard-spheres of diameter
$\sigma_{11}$ create a cavity which can incorporate a hard-sphere with
diameter up to $d_c=(2/\sqrt(3)-1)\sigma_{11}=0.1547\sigma_{11}$
making contact with each of the three spheres. Hence, if the geometric
condition $ \sigma_{12}^{NA}+\delta < (\sigma_{11} +d_c)/2$ is met, the
floating bond can only be simultaneously involved in two attractive
interactions.  
%%%%%{\bf FORSE la condizione sopra andrebbe spiegata meglio, magari con un disegno o forse riscritta come $\sigma_{12}^{NA}+\delta < (\sigma_{11}+d_c)/2$}
We choose $\sigma_{12}^{NA}=0.55\sigma_{11}$ and
$\delta=0.03\sigma_{12}^{NA}$ so that
$2(\sigma_{12}^{NA}+\delta)-\sigma_{11}=0.134\sigma_{11}< d_c$.  As a
result of this choice, a floating bond can be isolated (with potential
energy zero), bonded to one particle (potential energy $-u_0$) or
bonded to two distinct particles (potential energy $-2u_0$). Only in
the last case, the floating bond provides a link between the two
particles, which are thus considered bonded and belonging to the same
cluster.

To satisfy the limited valence condition (ii) we fix the hard-sphere
diameter of the floating bonds $\sigma_{22}$, determining the closest
distance between them.  To model a system with valence four (with
tetrahedral ordering) we choose $\sigma_{22}=0.8\sigma_{11}$ (and
hence, the mixture considered here has a negative non-additive
parameter $\Delta=\sigma_{12}^{NA}/(\sigma_{11}+\sigma_{22})-1 \simeq
-0.69$). This choice is dictated by geometric considerations.  Indeed,
the distance between vertices of a perfect tetrahedron is $2\sqrt{6}/3$ times
the distance between the center and the vertex.  
%{\bf QUI PER ME ERA CONFUSO. 
%IO NON VEDO IL MOTIVO DI PARLARE DI TETRAEDRO FORMATO DA
%PARTICELLE.  PREFERISCO PARLARE DI TETRAEDRO FORMATO DA UNA PARTICELLA
%DECORATA DA 4 FLOATING BONDS.}  
The distance of an interacting floating bond from the center of a
particle varies between $\sigma_{12}^{NA}$ and
$\sigma_{12}^{NA}+\delta$.  Hence, a tetrahedral arrangement requires
a value of $\sigma_{22}< 0.93 \sigma_{11}$.  To prevent the possibility of a
planar square arrangement of bonds (a geometry which would allow a
valence of six, with four bonds on the equatorial plane and two bonds
on the poles) we need to impose $\sigma_{22} > 0.802 \sigma_{11}$.  We
have checked that our choice $\sigma_{22}=0.8 \sigma_{11}$ never
generates particles connected to more than four floating bonds.
Finally we fix the number ratio between particles and floating bonds
following the stoichiometry of the system, i.e.  by imposing that in
the fully bonded ground state all particles participate to four bonds
and that all floating bonds have energy $-2u_0$., i.e. $N_2/N_1=2$.

We study a system of $N_1=1000$ particles and perform standard
event-driven MD simulations in the NVE ensemble.  For the smallest
studied temperatures, equilibration requires several months of CPU
time.  Packing fraction is defined as $\phi=\pi
N_1(\sigma_{11}/L)^3/6$, i.e. as the fraction of volume occupied only
by the particles.  In these units, taking into account the minimal and
maximal bond distance between two particles, the diamond crystal is
mechanically stable between $ 0.233 <
\phi < 0.255 $ (since a diamond crystal of touching hard-spheres has
$\phi\simeq0.340$). Units of length and energy are $\sigma_{11}$ and
$u_0$, while the Boltzmann constant $k_B$ is set equal to $1$.  We
equilibrated the mixture for a wide range of $\phi$ and $T$, up to
$\phi=0.56$ and down to $T=0.065$. We notice that only for $\phi
\gtrsim 0.52$ crystallization (into a fcc structure) was sometimes 
(but not always) detected at low $T$.

%\subsection{Bond definition and evaluation of connectivity properties}

The use of a square-well potential to model the interactions makes it
possible to unambiguously define the existence of a bond between two
large particle. Indeed, if the energy of a floating bonding particle
is $-2u_0$, then the two large particles interacting with the floating
bond particle are considered bonded.  In this way, large particles can
be partitioned into different clusters and the connectivity properties
of them can be examined. More explicitly, to test for percolation, the
simulation box is duplicated in all directions, and the ability of the
largest cluster to span the replicated system is controlled. If the
cluster in the simulation box does not connect with its copy in the
duplicated system, then the configuration is assumed to be
nonpercolating. The boundary between a percolating and a
nonpercolating state point is then defined as the probability of
observing infinite clusters in 50$\%$ of the configurations.

\section{Static properties}

\subsection{Radial distribution functions}
\begin{figure}
\includegraphics[width=.5\textwidth]{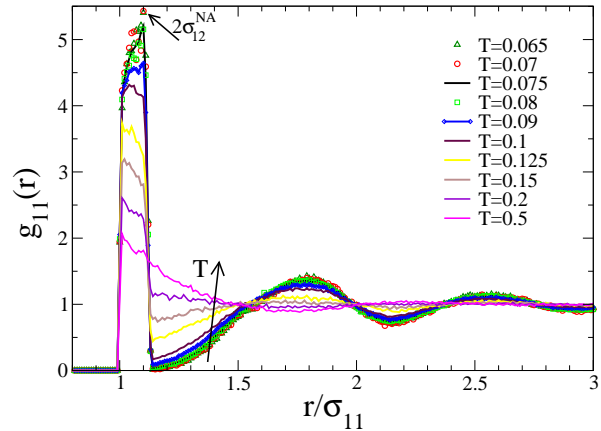}
\caption{Partial radial distribution function of particles $g_{11}(r)$  along a low-$\phi$ isochore, i.e. $\phi=0.25$, and upon decreasing $T$.}  
%COME MAI NON C'E' LA T=0.06 VISTO CHE DICIAMO CHE ABBIAMO STUDIATO ANCHE QUELLA? --- FORSE SAREBBERO DI AIUTO FRECCE CON SIMBOLI PER SIGMA11 e 2(SIGMA12+DELTA) --- IN TUTTE LE FIGURE SIMBOLI CIFRE PIU' GRANDI PLEASE}
\label{fig:gr-025}
\end{figure}

To visualize how directional interactions build up for large particles
upon decreasing temperature we report the behaviour of the particles
radial distribution function $g_{11}(r)$
% at fixed packing fraction $\phi=0.25$ and various $T$ 
in Fig.~\ref{fig:gr-025}. 
At large $T$, no significant correlations are present and the system behaves almost as a hard-sphere fluid mixture. The first peak of
$g_{11}(r)$ is found at $\sigma_{11}$. However, as $T$ is
lowered, more and more particles are bonded within the attractive
well, so that for
%$2\sigma_{12}^{NA} < r < 2(\sigma_{12}^{NA}+\delta)$, 
$\sigma_{11} < r <  2(\sigma_{12}^{NA}+\delta)$
a larger and larger correlation is built up. For 
$r>2(\sigma_{12}^{NA}+\delta)$ 
%$r> b_d^{MAX}$ 
an anti-correlation develops. When $T\lesssim 0.1$, the peak of
$g_{11}(r)$ shifts from $\sigma_{11}$ to $ 2\sigma_{12}^{NA}$, a
signature of the progressive role played by the floating bonds in
structuring the particles: while the growth at $\sigma_{11}$
saturates, that at $2\sigma_{12}^{NA}$ continues to increase.  
%% The
%% signal of the first peak is found up to $r \leq b_d^{MAX}$, while just
%% above this value $g_{11}(r)$ sharply decreases to almost identically
%% zero (since this distance becomes always occupied by floating bonds,
%% and thus unavailable to other particles).

A clear signal of the progressive structuring upon cooling is
provided by the increase of the nearest neighbors peaks.
The second peak is centered at a
distance $\approx 1.78\sigma_{11}$, 
%($\sim 1.62 b_d^{MIN}$), 
while
the third one is found at $\approx 2.53\sigma_{11}$.
% ($\sim 2.23 b_d^{MIN}$). 
Such a sequence of peak positions, with non-integer ratios, is typical
of network-forming systems with tetrahedral
arrangement\cite{Hor99aPRB,giovambattista,loertingnicolas}.  It is
also important to notice that, for $T<0.09$, such peaks do not show a
significant variation in intensity, indicating that the structuring
process is essentially completed and that the system has approached an
almost perfect long-range tetrahedral arrangement.

\begin{figure}
\includegraphics[width=.5\textwidth]{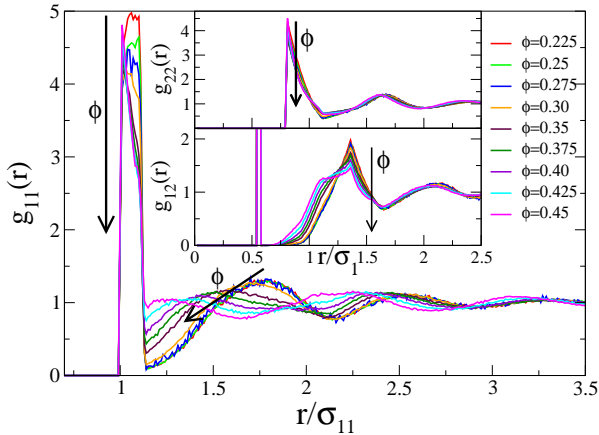}
\caption{Partial radial distribution functions of particles 
$g_{11}(r)$ at fixed low $T=0.09$ and varying $\phi$. In the insets
also partial radial distribution functions of floating bonds
$g_{22}(r)$ and of mixed type $g_{12}(r)$ are shown.
%A further magnification of the contact peak for $g_{12}(r)$ is also shown.  
}
\label{fig:gr-T009}
\end{figure}

The density dependence of the partial radial distribution functions is
reported, for $T=0.09$, in Fig.~\ref{fig:gr-T009}.  We first focus on
the evolution of $g_{11}(r)$ in the main panel. Data show that the
tetrahedral order, clearly visible at small and intermediate densities
($0.225<\phi<0.40$), is progressively lost with increasing $\phi$. At
this $T$, the curve corresponding to $\phi=0.30$ already shows small
deviations in the location of the second peak. For larger $\phi$ the
location of the second peak moves to smaller distances and its
amplitude decreases.  Now the first peak is always found at
$\sigma_{11}$ rather than $\sigma_{12}^{NA}$ for $\phi\gtrsim0.30$.  A
similar shift and rearrangement is observed for secondary peaks, as
well as for the evolution of the other partial distribution functions
$g_{12}(r)$ and $g_{22}(r)$, shown in the insets.

We further notice (not shown) that the $\phi$-value where tetrahedricity starts to
get progressively lost is found to depend on $T$.  On further
decreasing $T$, the maximum $\phi$ with dominant local order
increases: at $T=0.08, 0.075$, also $\phi=0.35$ becomes more
tetrahedral-like. However, for $\phi=0.40$, the structure even at the
lowest equilibrated $T$, which is actually the closest point to the
ground state structure that we could determine in our long-running
simulations (see below), remains always far from that of a tetrahedral
network.

These results suggest that geometric frustration due to packing acts
against the formation of a tetrahedral network above a certain $\phi$,
giving rise to a competition between energetic (directional
attraction) and entropic (excluded volume) interactions.

\subsection{Static structure factor}

We also study the behaviour of the normalized partial static structure
factors, i.e.  $S_{ij}(q)=\langle |\rho_{j}({\bf q}) \rho_{i}({\bf
q})^* |
\rangle/(N_i N_j)^{1/2}$, where $\rho_{j}({\bf q})=\sum_{m=1}^{N_j}
\exp{(i{\bf q}\cdot {\bf r}_m)}$ is the wave vector ${\bf q}$
component of the density of species $j$ (and the sum runs over all
$N_j$ particles of type $j$).  We focus only on the partial structure
factor of particles $S_{11}(q)$, aiming to emphasize those features
that are characteristic of the establishment of a fully connected
network of bonds.
\begin{figure}
\includegraphics[width=.5\textwidth]{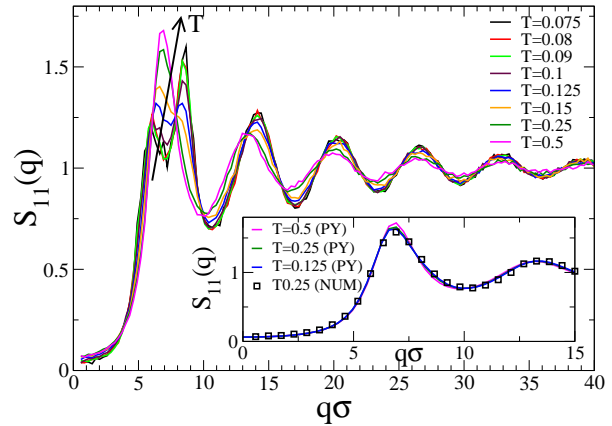}
\caption{ Normalized $S_{11}(q)$ for $\phi=0.35$ and varying $T$. 
Inset: Comparison with PY theoretical results.}
\label{fig:sq-0.35}
\end{figure}
Figure~\ref{fig:sq-0.35} shows the behavior of $S_{11}(q)$ as a
function of $T$ at $\phi=0.35$.  We notice that, at high $T$,
$S_{11}(q)$ displays a nearest-neighbour peak centered around
$q\sigma\approx 7$, characteristic of the hard-sphere interactions and
of all simple liquids. As $T$ is decreased the height of such a peak
also decreases, eventually giving rise at low $T$ to a splitting in
two peaks: one for larger length-scales at $q\sigma\sim 5$ and one at
smaller length-scales $q\sigma \sim 8.5$.  The first one seems to
saturate at low $T$, while the second continues to increase in
amplitude. The splitting of the main peak into two components is
characteristic of network forming tetrahedral liquids
\cite{Hor99aPRB,starrsq} and it is associated to the formation of an
open local structure.

Simple integral equation theories are not able to predict the
splitting of the peak on cooling. We solved numerically the binary
Ornstein-Zernike equation for our mixture within Percus-Yevick (PY)
approximation\cite{hansen06} at the same packing fraction $\phi=0.35$ upon lowering
$T$. Results are shown in the inset of Fig.~\ref{fig:sq-0.35},
together with a high-$T$ numerical curve.  While at high $T$ PY
provides a good description of the system, it fails severely upon
lowering $T$, where the splitting of the main peak is not at all
captured. Moreover, numerical convergence of the PY solution can not
be achieved for $T < 0.125$. Hence, the PY solution is only able to
capture a small decrease of the peak, without accounting for the
formation of the tetrahedral network, as expected due to the symmetric
approximation contained in PY.  This stresses the importance to
develop more elaborated integral equations which are able to account
for the angular correlations introduced by the bonding. In this
respect, a step forward could be made by  exploring  PY approximation applied
to the associative Ornstein-Zernike equation (PY-AOZ) which generally
provides better results, compared to simulations, for network-forming
liquids\cite{duda1,duda2}.

\begin{figure}
\includegraphics[width=.5\textwidth]{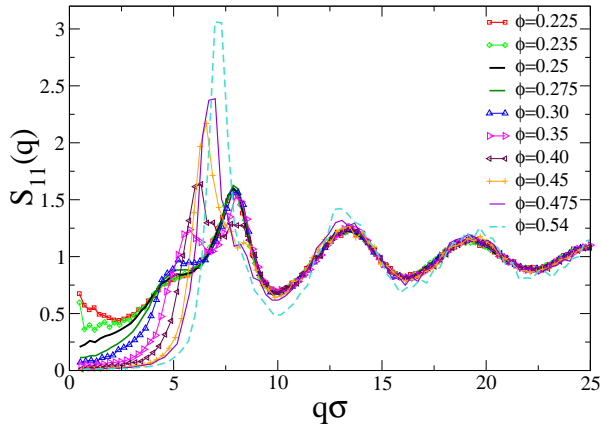}
\caption{Normalized $S_{11}(q)$ for  $T=0.09$ and varying $\phi$.}
\label{fig:sq-T009}
\end{figure}

The structure factor provides also information on the proximity to an
unstable region since the low $q$ behavior is related to the system
compressibility. Indeed, close to an unstable state of the particles,
$S_{11}(q)$ increases significantly at small $q$.  Figure
~\ref{fig:sq-T009} shows $S_{11}(q)$ as a function of packing fraction, from
$\phi=0.225$ (just close to the phase separation liquid boundary) to
$\phi=0.54$ along a low temperature isotherm ($T=0.09$).  On
increasing $\phi$, the critical low $q$ fluctuations disappear, giving
rise to the network two-peak structure, then crossing continuously to
a single-peaked standard $S_{11}(q)$ that finally resembles that of high
density simple liquids for $\phi > 0.40$.
% This evolution is similar to that observed in water\cite{starr,etc}.....

The study of the structural properties reported so far clearly
indicates that the possibility of forming a well-connected disordered
tetrahedral network arises only in a finite window of intermediate
densities,  i.e. $0.225 \lesssim
\phi \gtrsim 0.40 $. For smaller $\phi$ values, particles are not close
enough to form a fully bonded structure and phase separation into a gas
and a bonded liquid is preferred. For larger values,
the local density around each particle becomes more and more
incompatible with the open structure which is characteristic of
tetrahedral networks and packing becomes the leading driving mechanism
controlling the structure.

\subsection{Energy}

In the present model, attraction takes place only between particles
and floating bonds. Moreover, the interaction range is such that each
floating bond can interact simultaneously only with up to two distinct
particles.  This has the very important consequence that, for the
chosen $N_1/N_2$ ratio, the energetic ground state of the system is
known, being equal to two times the number of floating bonds (in units
of $-u_0$).  In the present model, this means that a fully bonded
configuration has a ground state energy $E_{gs}=- 2 u_0 N_2 $ or
equivalently $E_{gs}= - 4 u_0 N_1$. We notice that an energetic gain
is at hand both when a particle sticks to a floating bond (of
contribution $-u_0$) and when a true bond is established between two
particles (of contribution $-2u_0$). To differentiate between these
two situations, we use $E$ to indicate the potential energy of the
system (i.e. proportional the number of connections between a particle
and a floating bond), while we use $E_b$ to count the potential energy
associated to particle-particle bond contributions (i.e. proportional
to the number of floating bonds with potential energy $-2u_0$). Of
course, the two quantities tend to become identical as $T$ is lowered.
%(at least on low-intermediate $\phi$. 
We then define the bond probability as  $p_b\equiv (E_b-E_{gs})/E_{gs}$.

The possibility of knowing theoretically the ground state of the
system (a property shared with other limited valence
models~\cite{Moreno_05,MorenoJCP,DeMichele_05,DeMichele_06,Bianchi_06})
offers the possibility of unambiguously tracking the approach to the
fully bonded state and detect the range of densities where indeed the
system may reach continuously $E_{gs}$.  It has been previously
suggested~\cite{Moreno_05} that the ability to reach in equilibrium
almost fully bonded states is indeed a specific feature of network
forming liquids.

\begin{figure}
\includegraphics[width=.5\textwidth]{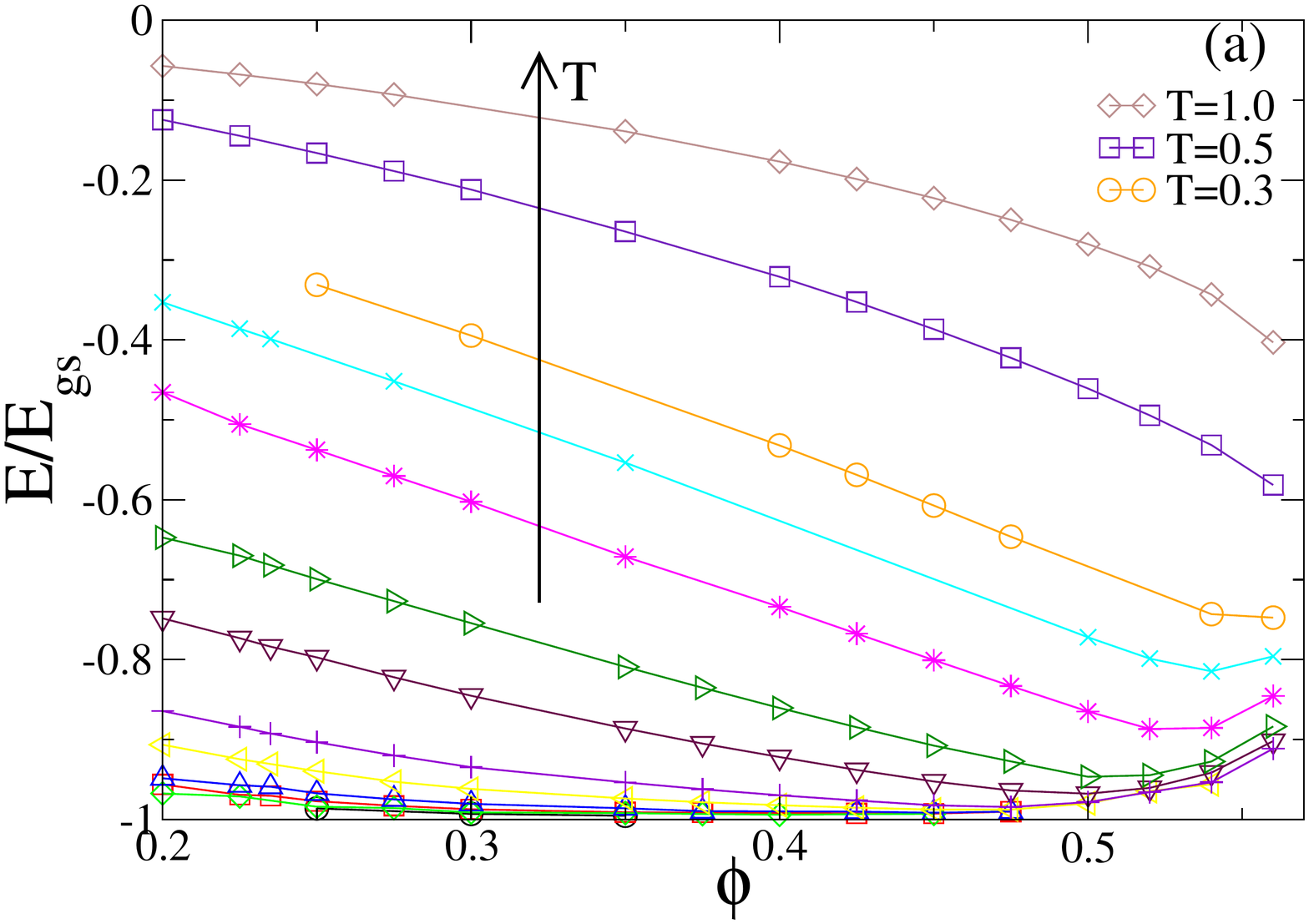}
\includegraphics[width=.5\textwidth]{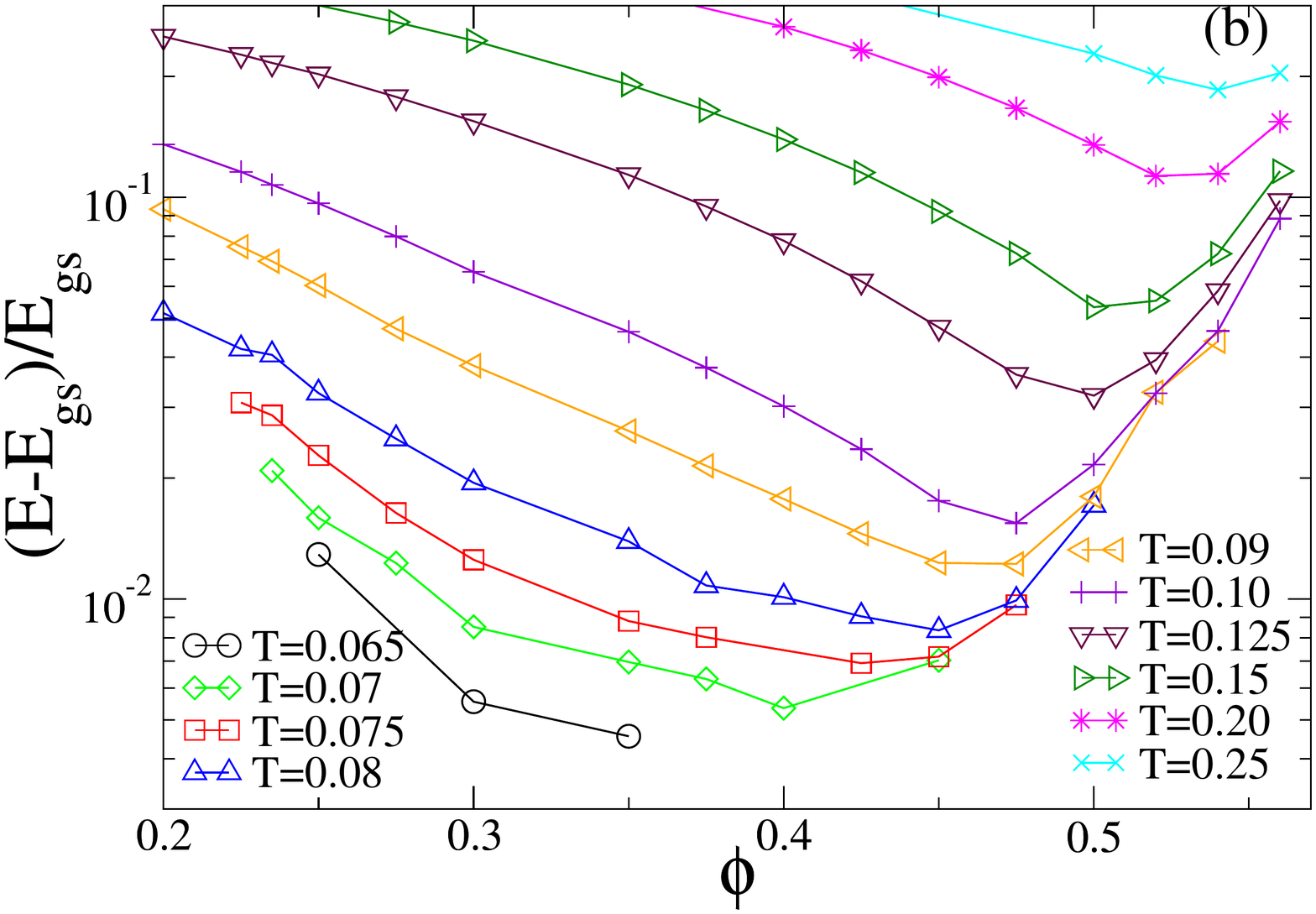}
\caption{Density dependence of the potential energy per particle relative to the ground state energy  along different isochores:
(a) $E/E_{gs}$ for all studied state points and (b)
$(E-E_{gs})/E_{gs}$ in a magnification of the low $T$ region in
semi-logarithmic scale.}
\label{fig:energy-T}
\end{figure}

Fig.~\ref{fig:energy-T}(a) shows the potential energy relative to the
ground state energy, expressed as $E/E_{gs}$, as a function of $\phi$
for all studied temperatures. A magnification in
semi-logarithmic scale of the low-$T$ isotherms is offered in
Fig.~\ref{fig:energy-T}(b) for the positive ratio $(E-E_{gs})/E_{gs}$.
%With this normalization, the negative of $E$ provides a 
%measure of the average number of bonds per large particle.  
Except for high temperatures, for which the $\phi$ dependence of $E$
is monotonic (as in simple square-well models), a minimum in density
appears. Since here the excluded volume interaction is modeled via a
hard-core interaction, the increase of $E$ at large $\phi$ can arise
only from a progressive breaking of the bonds.  For intermediate $T$,
when the average number of bonds per particle is still small, the
minimum is met only at very large $\phi$.  But, upon lowering $T$,
bonding becomes more and more extensive and the density at the minimum
progressively decreases, reaching $\phi \approx 0.3$ at the lowest
studied $T$. The $T$-dependence of the density minimum is shown in the
phase diagram reported below (see Fig. ~\ref{fig:phase}).
%Fig.~\ref{fig:min} .  COMMENTARE

Fig.~\ref{fig:energy-T}(b) also shows that, in a large window of
densities, the system is able to essentially reach the disordered
ground state. Indeed, more than 95$\%$ of the bonds (intended as
$E_b$) are formed.  This window of densities is roughly the same for
which structural properties suggested the presence of a tetrahedral
network of bonds. Since particles have already formed almost all
possible bonds at the lowest studied temperatures, a further decrease in $T$,
beyond the ones which we have been able to investigate, will not lead
to further significant structural changes.

\begin{figure}
\includegraphics[width=.5\textwidth]{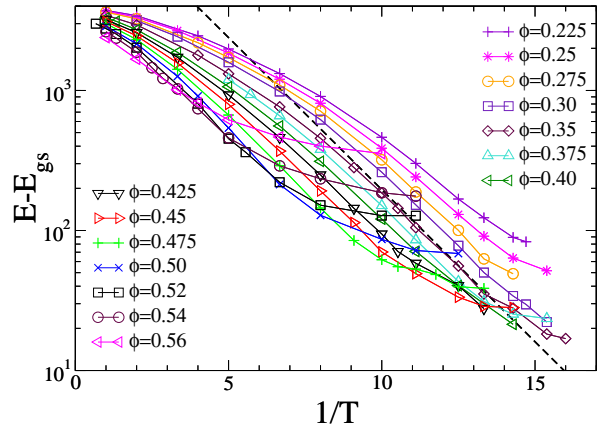}
\caption{Potential energy per particle $E-E_{gs}$ vs. $1/T$. 
Note that at intermediate densities an Arrhenius approach to the
ground state is observed. The dashed line is a reference curve with
activation energy $-0.5$.}
\label{fig:arrh-energy}
\end{figure}

Being the ground state energy a priori known, it is possible to
investigate the $T$-dependence of the energy on approaching the fully
bonded state. To this aim, Fig.~\ref{fig:arrh-energy} shows $E-E_{gs}$
vs $1/T$ in an Arrhenius plot for all investigated densities. Low
$\phi$ data are limited to the $T$-region above the liquid-gas
instability.  In the region where phase separation is not encountered,
a clear Arrhenius dependence is found. However, for large enough
densities, the energy decrease tends to saturate to a finite constant
value.  Only at intermediate densities ($0.225 \lesssim \phi\lesssim
0.40$), it appears to be possible to reach a fully bonded (defect
free) network. At larger densities the lowest energy state appears to
be larger than the fully bonded one, suggesting that the total volume
constraint imposes the presence of a finite number of defects in the
fully bonded network.  This is not surprising in view of the fact that
the establishment of a tetrahedrally coordinated environment enforces
a constraint on the local density.

\begin{figure}
\includegraphics[width=.5\textwidth]{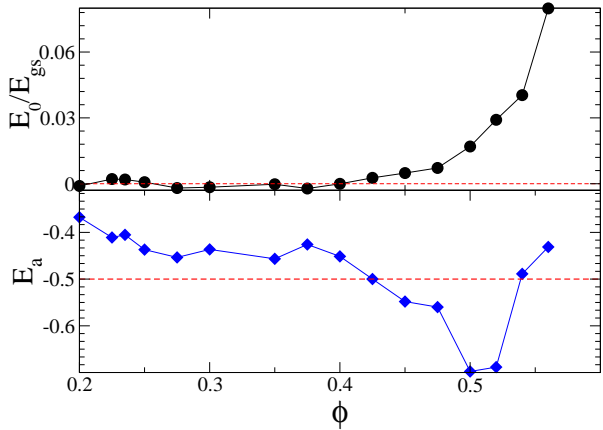}
\caption{Fit parameters from the fitting Arrhenius law in 
Eq. ~\protect\ref{eq:arrh}: effective ground-state energy ratio $E_0/E_{gs}$ (top) and
activation energy $E_a$ (bottom).}
\label{fig:arrhfit}
\end{figure}

To quantify the Arrhenius behaviour in an unbiased way, we fit the
$T$-dependence of the potential energy with the functional form 
\begin{equation}
[E-E_{gs}]= E_0+B \exp(E_a/T)
\label{eq:arrh}
\end{equation}
 in the region where a sufficient bonding is present. To realize this
condition, we consider only state points where $(E-E_{gs})/E_{gs} >
0.825$, in order to have the same fitting conditions for all studied
isochores.  With the chosen fitting function, we can fully reproduce
the behaviour of $E$ both in the region where a fully connected
network state is reached as well as that for states where frustration
comes into play.  Three fitting parameters are involved: $E_0$, $B$
and the activation energy $E_a$.  The results of the fits are
summarized in Fig.~\ref{fig:arrhfit} for all studied $\phi$. We
observe that the system approaches the expected ground state ($E_0=0$)
up to roughly $\phi\sim 0.40$, while a consistent increase of $E_0$ is
found at larger $\phi$. Hence, beyond such $\phi$ value, the system
can not reach a fully bonded configuration due to the excluded-volume
contributions in the free energy. Even at vanishing temperatures, when
minimization of the energy is the dominant driving force, it is
impossible for geometric reasons to reach a fully bonded state and the
effective ground state, reached along a constant $\phi$ cooling path,
is characterized by a non-zero fraction of unformed bonds.  Data in
Fig.~\ref{fig:arrh-energy} show that it is possible to
unambigously define an optimal region for network formation for
$0.235\lesssim \phi \lesssim 0.40$ (the lower bound being fixed by the
presence of phase separation).  In this density window, the fully
bonded ground state can be reached in equilibrium on isochoric
cooling.

It is interesting to observe that the
presence of a minimum in $E(\phi)$ along isotherms is consistent with
a defect-free ground state, but only in a small region of densities. The
optimal network-forming region largely coincides with that where the
structural properties also show tethrahedral arrangement, with the
exception of the state points close to $\phi=0.40$, where still
$E_0\rightarrow 0$ even though the network begins to be  deformed
due to the increase in packing.  Such feature is only possible in the
present model due to the flexibility of the chosen interaction parameters.
% (and in particular due to the gap between $\sigma_{22}$ and the perfect tethadrehon location distance).

The $\phi$-dependence of the activation energy (bottom panel) is also
instructive. It is observed that $E_a$ is slightly larger than $-0.5$
within the optimal network-forming region, showing an almost monotonic
decrease up to $\phi \sim 0.50$, then a reversal of trend is observed
at the highest studied values.  Note that the value $0.5$ is the
theoretical value expected for the breaking process of independent
bonds\cite{Werth1}.  

%{\bf The analysis of $1-p_b$ reveals a very similar Arrhenius dependence.
%UN PO APPESA --- LEVARE ?}

\begin{figure}
\includegraphics[width=.5\textwidth]{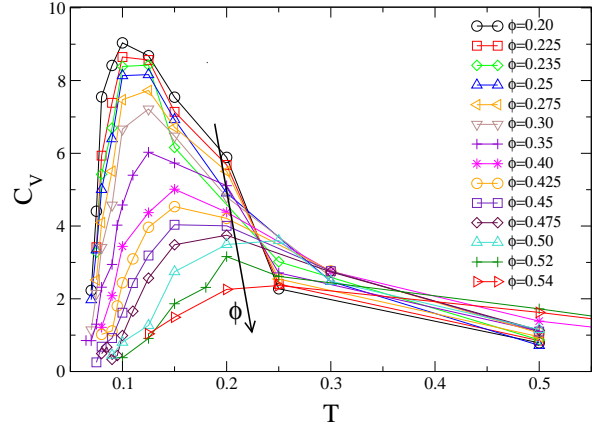}
\caption{Constant volume specific heat $C_V=(dE/dT)_V$ 
for all studied isochores.
}
\label{fig:cv}
\end{figure}

The low $T$ Arrhenius behavior carries with it another important
thermodynamic feature, the presence of a maximum in the constant
volume $V$ specific heat $C_V$.  The behaviour of $C_V$ along the
studied isochores is reported in Fig.~\ref{fig:cv}. In this respect,
the present model confirms the suggestion\cite{MorenoJCP} that a line
of $C_V$ maxima in the phase diagram is present when
bonding is the relevant driving force.  Differently from the model in
Ref.~\cite{MorenoJCP}, a monotonic increase of the $T$-value of the
$C_V$-maxima is observed upon increasing $\phi$. It is not a
coincidence that maxima in $C_V$ are also characteristic of reversible
self-assembly processes\cite{Sci07a,Bia07a}, i.e. of systems in which
bonding plays the leading role. The locus of $C_V$-maxima extracted
from these results  will be discussed in Sec.~\ref{sec:phase}.
%is also shown in the phase diagram of Fig. ~\ref{fig:phase}.

\subsection{Angular distribution of bonds}

\begin{figure}
\includegraphics[width=.5\textwidth]{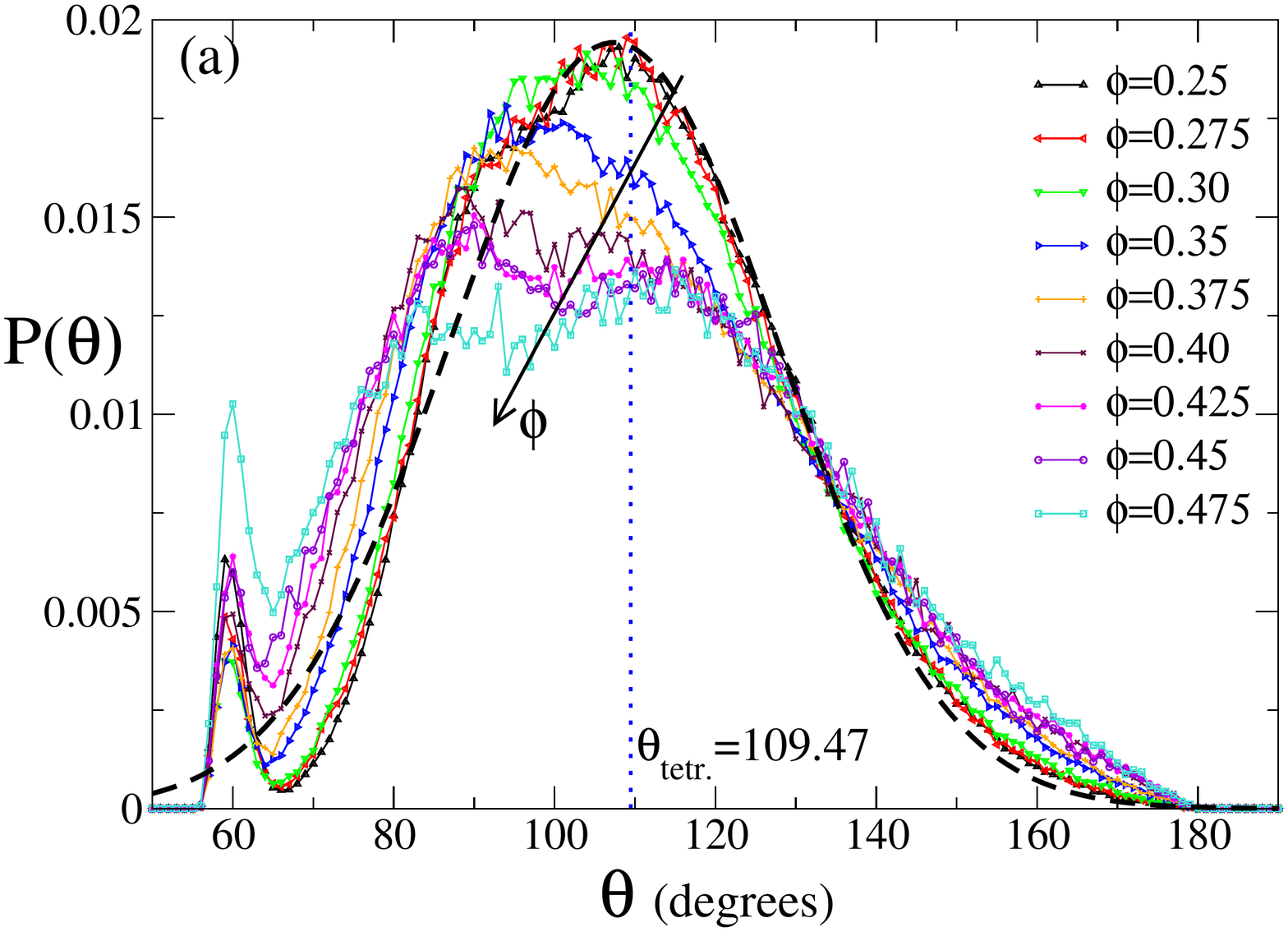}
\includegraphics[width=.5\textwidth]{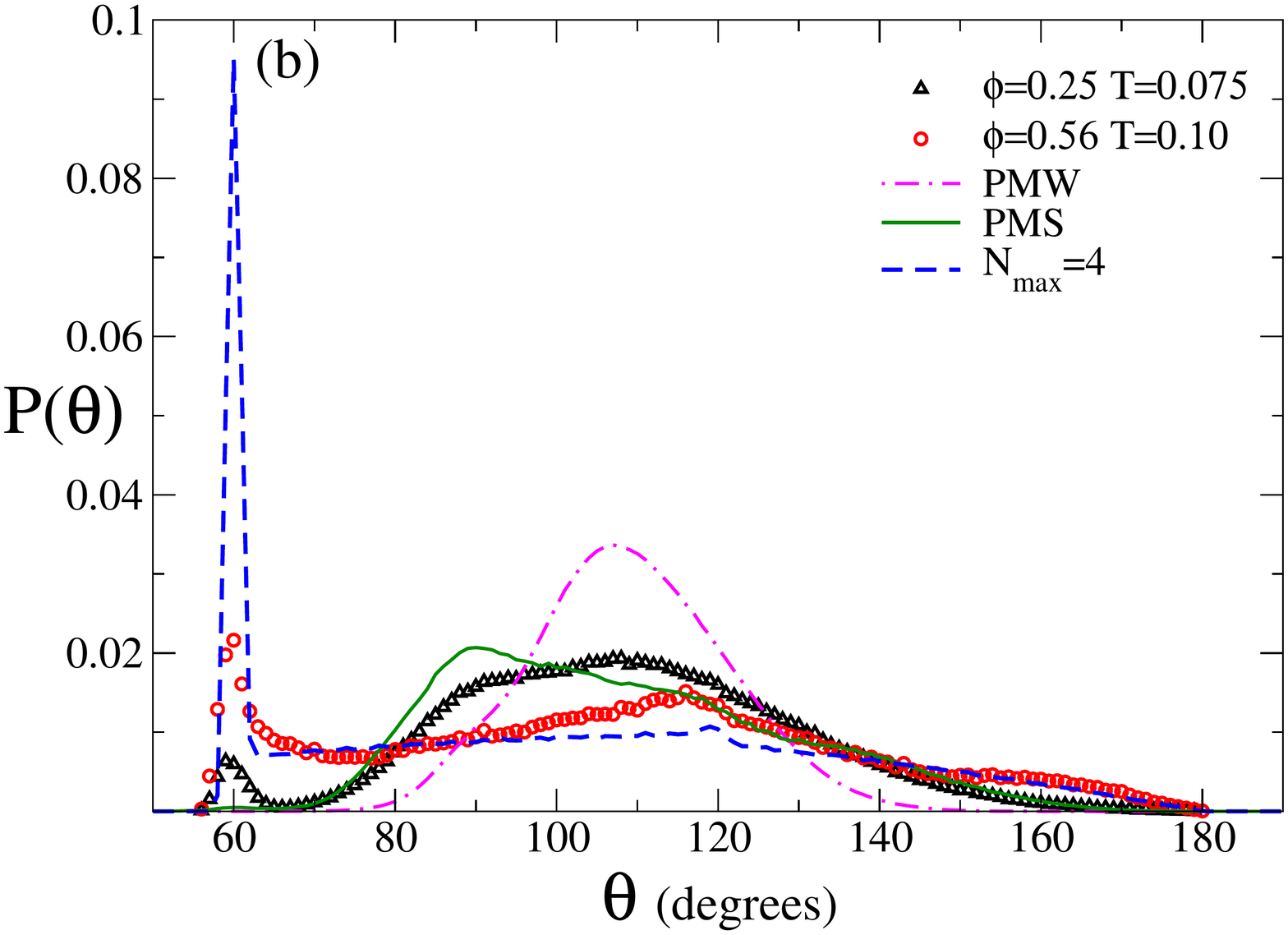}
\caption{
Distribution of the angle $\theta$ between bonded particles triplets.
Full thin lines are guides to the eye.  (a) $P(\theta)$ for various
$\phi$ at $T=0.075$. The thick dashed line is a Gaussian fit to the
low-$\phi$ states, suggesting an average angle $\sim 107.4$ and
variance $\sim 29.0$; (b) Comparison of $P(\theta)$ in  the
optimal network-forming density ($\phi=0.25, T=0.075$-- triangles )
with different 4-coordinated models: PWM (dot-dashed line), PSM (full
line), $N_{\rm max}$ (dashed line)  (all from Ref.~\protect\cite{DeMichele_06}), as well as for the present model at two different state points (triangles and circles).  }
\label{fig:angle}
\end{figure}

A direct measure of the tetrahedral structure of the network is offered by
the evaluation of the distribution of the angle $\theta$ between
bonded triplets of large particles, repeating the analysis that is
usually performed for similar models\cite{Rino93,VolPRB,Mou97a}.  The angle
distribution $P(\theta)$ is shown as a function of $\phi$ for
$T=0.075$ in Fig. ~\ref{fig:angle}(a). For $\phi
\lesssim 0.30$, $P(\theta)$ is almost independent of $\phi$,
suggesting the approach to a tetrahedral network with almost fully
connected particles. A peak close to the tetrahedral angle is found,
and an average value of the triplet angle of $\approx 108.2$. Also,
around $\theta=0.60$ a secondary peak is observed, a geometric
signature of the presence of a small number of three-particle
rings\cite{Rino93,VolPRB}. A Gaussian fit is attempted (dashed curve in
Fig. ~\ref{fig:angle}(a) ), which works well only in the
large-$\theta$ region, due to the presence of rings.  In agreement
with the structural indicators discussed previously,  $P(\theta)$
deviates more and more from the typical tetrahedral shape on
increasing $\phi$.  When the excluded volume contribution becomes
dominant over the attractive interactions, the main tetrahedral peak
is progressively lost and a wider distribution is observed, accompanied
also by a smaller average angle and by an increase of
three-particles loops.

In Fig. ~\ref{fig:angle}(b) a comparison of $P(\theta)$ for different
4-coordinated models is offered, completing the data already presented
in \cite{DeMichele_06}.  In that work, $P(\theta)$ for a model without
any geometric correlation in the location of the bonding sites (the
$N_{max}$ model\cite{Spe94,Spe96,Zac06JCP}) --- showing a rather flat distribution of
angles due to the lack of preferred orientation between the bonds
(i.e. random bond organization) --- was compared with $P(\theta)$ for
two models with well defined locations of the bonding sites on the
particles surface, the so-called primitive models of water
(PMW)\cite{Kol_87} and primitive models of silica
(PMS)\cite{Ford_04}. The PMW, at its optimal density, displays a quite
sharp and narrow peak, with an average angle $\simeq 109$ degrees, due
to the rigid organization of the bonding sites modeling the hydrogen
atoms and the lone-pairs of the water molecule, with a roughly Gaussian shape
of $P(\theta)$. On the other hand, the PMS at its optimal network-forming
density has a wider and more asymmetric distribution, which is peaked
at lower angles $\theta\approx 90$ degrees.  In addition, the
$N_{max}$ model was found to exhibit a dominant contribution of
three-particles loops or touching triplets ($\theta=60$ degrees), a
feature that was absent in PMW and negligible in PMS.  Comparing the
results for the present binary model to the existing data, we observe
that, within the optimal network-forming region, the organization of
particles is very reminiscent of that of PMS, except for a shift of
the peak position, which is closer to the tetrahedral position, as
well as for a more pronounced presence of loops. Interestingly, when
$\phi$ is increased, local tetrahedral order is lost, as for the
reported $\phi=0.56$-curve, and $P(\theta)$ tends to the $N_{max}$
distribution, confirming that the organization of the bonds becomes
increasingly random.

%Also, a Gaussian fit was attempted providing a slightly lower value. However, the fit does not work well because a significant fraction of touching triplets of large spheres is found at all $\phi$ (and also increasing with $\phi$), a feature renders the whole distributions always asymmetric on the low-angle side (hence a Gaussian form can only describe the large-angles). The three-particle loops were dominant for the $N_{max}$, absent for water, few for silica, and are non-negligible here, providing a measure of the flexibility of the large particles interactions. 

\subsection{Liquid-Gas unstable region}

Although we do not perform accurate phase coexistence studies, we can
provide a rough estimate of the location of the liquid-gas phase
separation region and of the associated critical point by a combined
check of the time evolution of the large length-scale structural
properties, such as $S(q\rightarrow 0)$, as well as of the pressure
behaviour along different isotherms.  The lowest investigated $\phi$
where phase separation can be ruled out at all studied temperatures is
$\phi=0.25$.
%Thus the liquid boundary of the spinodal line is located in the region $0.235 \leq \phi \leq 0.25$.  
For lower $\phi$, we detect an increase of $S(q\rightarrow 0)$ as well
as the development of maxima and minima in the behaviour of $P(\phi)$
(coming respectively from small and large $\phi$). We define, for each
$T$, the location in $\phi$ of such maxima and minima as the spinodal
points.  We note that at $T=0.1$ the pressure dependence on $\phi$
does not show any loop, while at $T=0.095$ a small instability region
seems to be present (within our numerical resolution) close to
$\phi\sim 0.10$.  These two temperatures thus bracket the critical
temperature $T_c$. Hence, we estimate $T_c=0.095\pm 0.005$, while
$\phi_c=0.10\pm 0.025$.  A more accurate evaluation of the critical
parameters, with appropriate techniques should be undertaken in future
studies. We also note that the value of $\phi$ at which no critical
fluctuations are observed in the entire investigated $T$-range is in
close agreement both with the $N_{\rm max}$ model\cite{Zaccagel} with
coordination number four as well as with the PMW\cite{DeMichele_05}
and PMS\cite{DeMichele_06} models, reinforcing the view that the
valency is the control parameter for the location of the liquid-gas
spinodal\cite{Bianchi_06}, under single-bond conditions.

\section{A global look at the phase diagram}
\label{sec:phase}
To summarize the static properties and to provide a coherent picture
of the system behaviour, we discuss here  the phase
diagram of the model,  in combination with other relevant thermodynamic and connectivity loci.
\begin{figure}
\includegraphics[width=.5\textwidth]{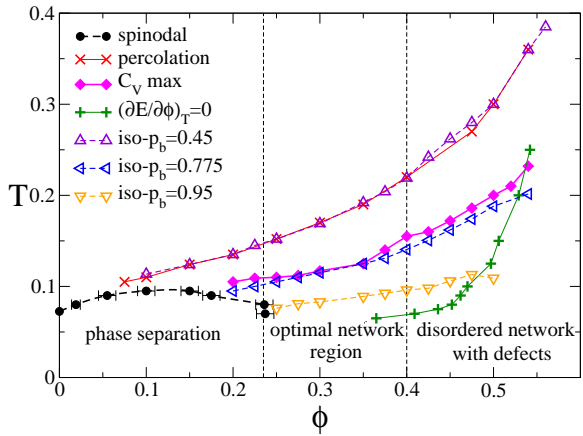}
\caption{Static phase diagram in the $(\phi,T)$ plane of the 
studied spherical binary model. The reported lines are: gas-liquid
spinodal (circles), percolation line (x-symbols), locus of $C_V$
maxima (diamonds), iso-$p_b$ lines at selected values (triangles) and
energy minima (crosses). Lines are guides to the eye. Vertical dashed
lines show the liquid boundary of the spinodal line ($\phi\sim 0.235$)
and the boundary ($\phi \sim 0.40$) between the optimal network
region, where bonds are primarily directional, and that of a
disordered network with defects, where directionality is progressively
lost.  }
\label{fig:phase}
\end{figure}

Fig.~\ref{fig:phase} shows the liquid-gas spinodal, the locus
encompassing the region of thermodynamic instability.  It is worth
emphasizing that such instability region is confined to densities
much smaller than those observed in spherical attractive potentials.
Liquid densities are roughly half of the typical value of normal
liquids, highlighting the large empty spaces which
characterize the formation of a tetrahedral network.  At the same
time, the small-$\phi$ location of the liquid branch of the spinodal
shows that, in this model, it is possible to find a region of
intermediate densities where, on lowering $T$, the system remains
homogeneous down to very small $T$, a feature which is not possible
with spherically symmetric attractive potentials. Moreover, the liquid
branch of the spinodal terminates in a region which is compatible with
the window of stability of the diamond crystal\cite{Monson_98,romanoJPCM}. This topology of the
liquid-gas coexistence is typical of models with directional
interactions, and it is here confirmed through the use of our
spherical binary mixture.

To provide an indication of the degree of bonding observed in
different regions of the phase diagram, Fig.~\ref{fig:phase} also
shows constant bond-probability lines (iso-$p_b$ lines), i.e. lines of
constant number of bonds per particles.  Here a bond is defined as a
floating bond with potential energy $-2u_0$. The lines have positive
slope, signaling that on increasing density bond formation becomes
favored. Only at very large $\phi$ and very large $p_b$ (only visible
for $p_b=0.95$ in the figure for the highest density point), a sharp
change of slope (flat, then negative at even larger $p_b$) is found,
associated to the disruptive effect of increasing density beyond the
values for optimal network formation.

Fig.~\ref{fig:phase} also shows the percolation line and the locus of
$C_V$ maxima.  State points on the right-hand side of the percolation
line are characterized by the presence of an infinite cluster in
$>50\%$ of the configurations. This definition of percolation locus is
strictly a geometric measure and does not provide any information on
the lifetime of the spanning cluster\cite{Zac07a}.  The percolation
locus coincides, within numerical resolution, with the $p_b=0.45$
locus, suggesting that a constant fraction of bonds is requested to
percolate independently of $\phi$. The $p_b$ value at percolation,
$p_b^{perc}\sim 0.45$, is slightly larger than the one for random bond
percolation on a diamond lattice, known to be
$p_b=0.388$\cite{Sta92book}, an effect which can be attributed to the
disordered distribution of particles in the fluid.  The percolation
locus is located above the critical point, confirming that the
development of long range correlated critical fluctuations requires as
a pre-requisite the existence of a spanning network of bonded
particles\cite{coniglio-klein}.  The locus of $C_v$ maxima is located
well inside the percolation region and, by comparing with the
iso-$p_b$ curves, it takes place when the bond probability is $\approx
0.75$.  Comparing with published data of models with different
valency\cite{Sci07a,Bia07a,MorenoJCP} it appears that the $C_V$ maxima
line progressively moves to larger and larger $p_b$ values with
increasing valence. The trend is thus opposite to the one
characterizing the valence dependence of $p_b^{perc}$.  Finally,
Fig.~\ref{fig:phase} also shows the locus of potential energy minima
along isotherms, i.e. for which $(\partial E/\partial\phi)_T=0$, a
line marking the crossover from an unconstrained network of bonds to
a state in which bonding can not be any longer optimized due to volume
constraints.

\begin{figure}
\includegraphics[width=.5\textwidth]{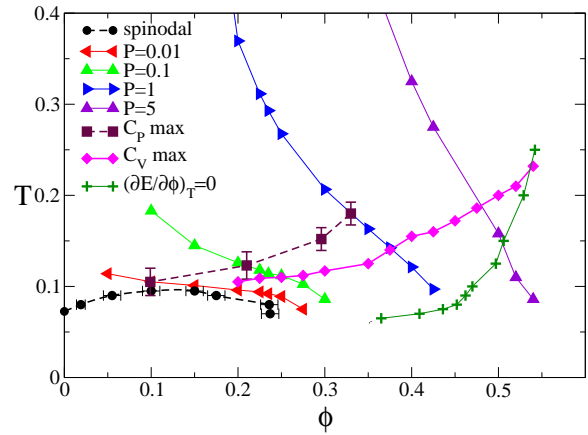}
\caption{Static phase diagram in the $(\phi,T)$ plane of the 
studied spherical binary model. In addition to lines reported in
Fig. ~\protect\ref{fig:phase}, i.e. gas-liquid spinodal (circles),
locus of $C_V$ maxima (diamonds) and of energy minima (crosses),
several isobaric paths are shown (triangles). Moreover, the locus
of $C_P$ maxima (squares) is also reported, merging into the spinodal
line at the critical point. }
\label{fig:phase2}
\end{figure}

Additional information on the $T$ and $\rho$ dependence of the
properties of the model are reported in Fig.~\ref{fig:phase2}. This
shows, in addition to some of the lines reported in the previous
figure, the location of selected isobars with the pressure $P$ varying
by more than two orders of magniude. From the constant-$P$ paths, we
evaluate the enthalpy $H=E+PV$, and we extract the behaviour of the
constant-pressure specific heat $C_P=(\partial H/\partial T)_P$. We find the presence
of clear maxima also in $C_P$, whose locus is also reported in
Fig. ~\ref{fig:phase2}. The line of $C_P$ maxima is an example of a
Widom line\cite{ScioPNAS,pradeepPNAS}, which starts, by definition,
from the critical point. Its calculation therefore also confirms the
previously discussed estimate of the critical point within our
numerical resolution.  Finally we observe that, in the investigated
$T$ region, there is no evidence of anomalies in the density (constant
$P$ density maxima), that are commonly found in water
models\cite{Poo92,st2jpcm,DeMichele_05}.

%The slope of these curves is a measure of $dT/d\phi_{P} \sim - \alpha^{-1}$, where $\alpha$ is the  thermal expansion coefficient. At low $T$,  the absolute value of the slope progressively decreases on lowering $P$.

%Such static picture will be complemented by an evaluation of the dynamical properties, and their interplay with thermodynamic behaviour, in a subsequent study.

\section{Conclusions}

In this article we have introduced an opportunely designed, spherical
binary mixture model, which is able to generate a fluid with a
effective coordination number of four.  Differently from previously
studied four-coordinate
model\cite{Zaccagel,Zac06JCP,DeMichele_05,DeMichele_06,Bianchi_06},
the particle-particle interactions are spherical, a feature which is
important in order to allow a future comparison with theoretical
approaches.

We have shown that the static phase diagram of the system is very
similar to the one reported for patchy
models\cite{Zac06JCP,DeMichele_05,DeMichele_06,Largo_07}, as well as
for more sophisticated models of network forming
liquids\cite{st2jpcm,bks}. Indeed, the unstable phase-separating
region of the system is confined to low packing fraction $\phi$ and
small $T$, consistently with
previous studies and Wertheim theory calculations \cite{Bianchi_06}
for four-coordinated particles. This reinforces the statement, that,
from the point of view of suppressing phase separation, the
arrangement of the sticky points onto the particle surface is not
qualitatively, but only quantitatively relevant.

The unstable gas-liquid region is followed at larger densities by an
optimal network region, i.e. a window of densities where the system
can be equilibrated down to very low temperatures observing a progressive
formation of a tetrahedral network of bonds.  In this region, the
system almost reaches the fully bonded configuration, i.e. the
disordered ground state of the system and hence a further lowering of
the temperature would not produce any significant structural change.
For even larger densities, we observe a destruction of the tetrahedral
bonding, induced by the packing constraints.  A signature of this
effect is observed in both the structure factor of the system, which
crosses from the tetrahedral-network form to the hard-sphere form as
well as in the progressive breaking of the bonds on isothermal
increase of the density.

It is also interesting to discuss the relative location of the
percolation locus, the locus of specific heat maxima and the
liquid-gas spinodal.  In the present model, the percolation line
provides the first indication of the clustering process upon lowering
$T$. Successive cooling brings to the presence of a line of $C_V$
maxima, and finally of the spinodal line.  The constant-volume
specific heat maxima line appears to be a characteristic of all bonded
systems, since it is observed also in the case of valence
two\cite{Sci07a} where no percolation is present. Its relative
location has an opposite dependence on the bond probability as
compared to the percolation line, when studied as a function of the
valence.  This suggests that, for values of the valence larger than
the one studied here, the line of $C_V$ maxima may become buried in the
region where equilibration is not feasible any longer, i.e. in the
glass state. This could be the reason why it is not observed in
standard spherical models, such as a simple square-well potential.

Finally, we note that the possibility of generating the complex
pattern characteristic of tetrahedral liquids with a simple binary
mixture with pair-wise additive square-well interactions opens the
possibility of applying the tools of modern liquid theory (which are
nowadays particularly accurate for spherical potentials) to the
present case.  It will also be possible, using theoretical and/or
numerical structure factors investigate the dynamics of the present
system using the formalism of the Mode Coupling Theory for the glass
transition. Work in this direction is underway and a companion paper
on the dynamics will appear shortly.

%The advantages of the model are multiple.  The tunability of the parameters allows for studying various geometrical situations, in connection for example with experiments. If the spots need not to be regularly distributed on the surface, different kinds of spots can be introduced. If more than one species of particles are involved, to each of them we can assign the corresponding ad-hoc sticky spots, balancing the stoichiometric ratio for the purposes of the study.  

We acknowledge support from MIUR Prin and MRTN-CT-2003-504712.
We thank C. N. Likos and I. Saika-Voivod for useful discussions.

\bibliographystyle{./apsrev}
\bibliography{./articoli,./altra,./advances,./biblio_patchy.bib,./star-star}

\begin{thebibliography}{58}
\expandafter\ifx\csname natexlab\endcsname\relax\def\natexlab#1{#1}\fi
\expandafter\ifx\csname bibnamefont\endcsname\relax
  \def\bibnamefont#1{#1}\fi
\expandafter\ifx\csname bibfnamefont\endcsname\relax
  \def\bibfnamefont#1{#1}\fi
\expandafter\ifx\csname citenamefont\endcsname\relax
  \def\citenamefont#1{#1}\fi
\expandafter\ifx\csname url\endcsname\relax
  \def\url#1{\texttt{#1}}\fi
\expandafter\ifx\csname urlprefix\endcsname\relax\def\urlprefix{URL }\fi
\providecommand{\bibinfo}[2]{#2}
\providecommand{\eprint}[2][]{\url{#2}}

\bibitem[{\citenamefont{Binder and Kob}(2005)}]{kobbook}
\bibinfo{author}{\bibfnamefont{K.}~\bibnamefont{Binder}} \bibnamefont{and}
  \bibinfo{author}{\bibfnamefont{W.}~\bibnamefont{Kob}},
  \emph{\bibinfo{title}{Glassy Materials and Disordered Solids: An Introduction
  to their Statistical Mechanics}} (\bibinfo{publisher}{World Scientific},
  \bibinfo{year}{2005}).

\bibitem[{\citenamefont{Angell}(1985)}]{Ang85a}
\bibinfo{author}{\bibfnamefont{C.~A.} \bibnamefont{Angell}},
  \bibinfo{journal}{J. Non-Cryst. Solids} \textbf{\bibinfo{volume}{73}},
  \bibinfo{pages}{1} (\bibinfo{year}{1985}).

\bibitem[{\citenamefont{G\"otze}(1991)}]{goetze}
\bibinfo{author}{\bibfnamefont{W.}~\bibnamefont{G\"otze}},
  \emph{\bibinfo{title}{Liquids, Freezing and the Glass Transition}}
  (\bibinfo{publisher}{North-Holland Amsterdam}, \bibinfo{year}{1991}), pp.
  \bibinfo{pages}{287--503}.

\bibitem[{\citenamefont{{Sciortino} and {Tartaglia}}(2005)}]{advances}
\bibinfo{author}{\bibfnamefont{F.}~\bibnamefont{{Sciortino}}} \bibnamefont{and}
  \bibinfo{author}{\bibfnamefont{P.}~\bibnamefont{{Tartaglia}}},
  \bibinfo{journal}{Advances in Physics} \textbf{\bibinfo{volume}{54}},
  \bibinfo{pages}{471} (\bibinfo{year}{2005}).

\bibitem[{\citenamefont{Zaccarelli}(2007)}]{Zac07a}
\bibinfo{author}{\bibfnamefont{E.}~\bibnamefont{Zaccarelli}},
  \bibinfo{journal}{J. Phys.: Condens. Matter} \textbf{\bibinfo{volume}{19}},
  \bibinfo{pages}{323101} (\bibinfo{year}{2007}).

\bibitem[{\citenamefont{{Voigtmann} and {Poon}}(2006)}]{Voigt06}
\bibinfo{author}{\bibfnamefont{T.}~\bibnamefont{{Voigtmann}}} \bibnamefont{and}
  \bibinfo{author}{\bibfnamefont{W.~C.~K.} \bibnamefont{{Poon}}},
  \bibinfo{journal}{J. Phys.: Condens. Matter} \textbf{\bibinfo{volume}{18}},
  \bibinfo{pages}{L465} (\bibinfo{year}{2006}).

\bibitem[{\citenamefont{{Sastry}}(2000)}]{Sas00PRL}
\bibinfo{author}{\bibfnamefont{S.}~\bibnamefont{{Sastry}}},
  \bibinfo{journal}{Phys. Rev. Lett.} \textbf{\bibinfo{volume}{85}},
  \bibinfo{pages}{590} (\bibinfo{year}{2000}).

\bibitem[{\citenamefont{Zaccarelli et~al.}(2004)\citenamefont{Zaccarelli,
  Sciortino, Buldyrev, and Tartaglia}}]{Zaccapri}
\bibinfo{author}{\bibfnamefont{E.}~\bibnamefont{Zaccarelli}},
  \bibinfo{author}{\bibfnamefont{F.}~\bibnamefont{Sciortino}},
  \bibinfo{author}{\bibfnamefont{S.~V.} \bibnamefont{Buldyrev}},
  \bibnamefont{and}
  \bibinfo{author}{\bibfnamefont{P.}~\bibnamefont{Tartaglia}},
  \emph{\bibinfo{title}{Short-ranged attractive colloids: What is the gel
  state?}} (\bibinfo{publisher}{Elsevier, Amsterdam}, \bibinfo{year}{2004}),
  pp. \bibinfo{pages}{181--194}.

\bibitem[{\citenamefont{{Foffi} et~al.}(2005)\citenamefont{{Foffi}, {De
  Michele}, {Sciortino}, and {Tartaglia}}}]{Fof05a}
\bibinfo{author}{\bibfnamefont{G.}~\bibnamefont{{Foffi}}},
  \bibinfo{author}{\bibfnamefont{C.}~\bibnamefont{{De Michele}}},
  \bibinfo{author}{\bibfnamefont{F.}~\bibnamefont{{Sciortino}}},
  \bibnamefont{and}
  \bibinfo{author}{\bibfnamefont{P.}~\bibnamefont{{Tartaglia}}},
  \bibinfo{journal}{Phys. Rev. Lett.} \textbf{\bibinfo{volume}{94}},
  \bibinfo{pages}{078301} (\bibinfo{year}{2005}).

\bibitem[{\citenamefont{Zaccarelli et~al.}(2005)\citenamefont{Zaccarelli,
  Buldyrev, {La Nave}, Moreno, Saika-Voivod, Sciortino, and
  Tartaglia}}]{Zaccagel}
\bibinfo{author}{\bibfnamefont{E.}~\bibnamefont{Zaccarelli}},
  \bibinfo{author}{\bibfnamefont{S.~V.} \bibnamefont{Buldyrev}},
  \bibinfo{author}{\bibfnamefont{E.}~\bibnamefont{{La Nave}}},
  \bibinfo{author}{\bibfnamefont{A.~J.} \bibnamefont{Moreno}},
  \bibinfo{author}{\bibfnamefont{I.}~\bibnamefont{Saika-Voivod}},
  \bibinfo{author}{\bibfnamefont{F.}~\bibnamefont{Sciortino}},
  \bibnamefont{and}
  \bibinfo{author}{\bibfnamefont{P.}~\bibnamefont{Tartaglia}},
  \bibinfo{journal}{Phys. Rev. Lett.} \textbf{\bibinfo{volume}{94}},
  \bibinfo{pages}{218301} (\bibinfo{year}{2005}).

\bibitem[{\citenamefont{{Bianchi} et~al.}(2006)\citenamefont{{Bianchi},
  {Largo}, {Tartaglia}, {Zaccarelli}, and {Sciortino}}}]{Bianchi_06}
\bibinfo{author}{\bibfnamefont{E.}~\bibnamefont{{Bianchi}}},
  \bibinfo{author}{\bibfnamefont{J.}~\bibnamefont{{Largo}}},
  \bibinfo{author}{\bibfnamefont{P.}~\bibnamefont{{Tartaglia}}},
  \bibinfo{author}{\bibfnamefont{E.}~\bibnamefont{{Zaccarelli}}},
  \bibnamefont{and}
  \bibinfo{author}{\bibfnamefont{F.}~\bibnamefont{{Sciortino}}},
  \bibinfo{journal}{Phys. Rev. Lett.} \textbf{\bibinfo{volume}{97}},
  \bibinfo{pages}{168301} (\bibinfo{year}{2006}).

\bibitem[{\citenamefont{Zaccarelli et~al.}(2006)\citenamefont{Zaccarelli,
  Saika-Voivod, Moreno, Buldyrev, Tartaglia, and Sciortino}}]{Zac06JCP}
\bibinfo{author}{\bibfnamefont{E.}~\bibnamefont{Zaccarelli}},
  \bibinfo{author}{\bibfnamefont{I.}~\bibnamefont{Saika-Voivod}},
  \bibinfo{author}{\bibfnamefont{A.~J.} \bibnamefont{Moreno}},
  \bibinfo{author}{\bibfnamefont{S.~V.} \bibnamefont{Buldyrev}},
  \bibinfo{author}{\bibfnamefont{P.}~\bibnamefont{Tartaglia}},
  \bibnamefont{and}
  \bibinfo{author}{\bibfnamefont{F.}~\bibnamefont{Sciortino}},
  \bibinfo{journal}{J. Chem. Phys.} \textbf{\bibinfo{volume}{{\bf 124}}},
  \bibinfo{pages}{124908} (\bibinfo{year}{2006}).

\bibitem[{\citenamefont{{De Michele}
  et~al.}(2006{\natexlab{a}})\citenamefont{{De Michele}, Gabrielli, Tartaglia,
  and Sciortino}}]{DeMichele_05}
\bibinfo{author}{\bibfnamefont{C.}~\bibnamefont{{De Michele}}},
  \bibinfo{author}{\bibfnamefont{S.}~\bibnamefont{Gabrielli}},
  \bibinfo{author}{\bibfnamefont{P.}~\bibnamefont{Tartaglia}},
  \bibnamefont{and}
  \bibinfo{author}{\bibfnamefont{F.}~\bibnamefont{Sciortino}},
  \bibinfo{journal}{J. Phys. Chem. B} \textbf{\bibinfo{volume}{110}},
  \bibinfo{pages}{8064} (\bibinfo{year}{2006}{\natexlab{a}}).

\bibitem[{\citenamefont{{De Michele}
  et~al.}(2006{\natexlab{b}})\citenamefont{{De Michele}, {Tartaglia}, and
  {Sciortino}}}]{DeMichele_06}
\bibinfo{author}{\bibfnamefont{C.}~\bibnamefont{{De Michele}}},
  \bibinfo{author}{\bibfnamefont{P.}~\bibnamefont{{Tartaglia}}},
  \bibnamefont{and}
  \bibinfo{author}{\bibfnamefont{F.}~\bibnamefont{{Sciortino}}},
  \bibinfo{journal}{J. Chem. Phys.} \textbf{\bibinfo{volume}{125}},
  \bibinfo{pages}{4710} (\bibinfo{year}{2006}{\natexlab{b}}).

\bibitem[{\citenamefont{{Schilling} and {Scheidsteger}}(1997)}]{Schi97a}
\bibinfo{author}{\bibfnamefont{R.}~\bibnamefont{{Schilling}}} \bibnamefont{and}
  \bibinfo{author}{\bibfnamefont{T.}~\bibnamefont{{Scheidsteger}}},
  \bibinfo{journal}{Phys. Rev. E} \textbf{\bibinfo{volume}{56}},
  \bibinfo{pages}{2932} (\bibinfo{year}{1997}).

\bibitem[{\citenamefont{Fabbian et~al.}(1999)\citenamefont{Fabbian, Latz,
  Schilling, Sciortino, Tartaglia, and Theis}}]{Fab99b}
\bibinfo{author}{\bibfnamefont{L.}~\bibnamefont{Fabbian}},
  \bibinfo{author}{\bibfnamefont{A.}~\bibnamefont{Latz}},
  \bibinfo{author}{\bibfnamefont{R.}~\bibnamefont{Schilling}},
  \bibinfo{author}{\bibfnamefont{F.}~\bibnamefont{Sciortino}},
  \bibinfo{author}{\bibfnamefont{P.}~\bibnamefont{Tartaglia}},
  \bibnamefont{and} \bibinfo{author}{\bibfnamefont{C.}~\bibnamefont{Theis}},
  \bibinfo{journal}{Phys. Rev. E} \textbf{\bibinfo{volume}{60}},
  \bibinfo{pages}{5768} (\bibinfo{year}{1999}).

\bibitem[{\citenamefont{{Schilling}}(2000)}]{Schi00b}
\bibinfo{author}{\bibfnamefont{R.}~\bibnamefont{{Schilling}}},
  \bibinfo{journal}{J. Phys.: Condens. Matter} \textbf{\bibinfo{volume}{12}},
  \bibinfo{pages}{6311} (\bibinfo{year}{2000}).

\bibitem[{\citenamefont{{Theis} et~al.}(2000)\citenamefont{{Theis},
  {Sciortino}, {Latz}, {Schilling}, and {Tartaglia}}}]{Schi00a}
\bibinfo{author}{\bibfnamefont{C.}~\bibnamefont{{Theis}}},
  \bibinfo{author}{\bibfnamefont{F.}~\bibnamefont{{Sciortino}}},
  \bibinfo{author}{\bibfnamefont{A.}~\bibnamefont{{Latz}}},
  \bibinfo{author}{\bibfnamefont{R.}~\bibnamefont{{Schilling}}},
  \bibnamefont{and}
  \bibinfo{author}{\bibfnamefont{P.}~\bibnamefont{{Tartaglia}}},
  \bibinfo{journal}{Phys. Rev. E} \textbf{\bibinfo{volume}{62}},
  \bibinfo{pages}{1856} (\bibinfo{year}{2000}).

\bibitem[{\citenamefont{Chong and Hirata}(1998)}]{Cho98a}
\bibinfo{author}{\bibfnamefont{S.~H.} \bibnamefont{Chong}} \bibnamefont{and}
  \bibinfo{author}{\bibfnamefont{F.}~\bibnamefont{Hirata}},
  \bibinfo{journal}{Phys. Rev. E} \textbf{\bibinfo{volume}{58}},
  \bibinfo{pages}{6188} (\bibinfo{year}{1998}).

\bibitem[{\citenamefont{Chong and Sciortino}(2004)}]{Cho04a}
\bibinfo{author}{\bibfnamefont{S.~H.} \bibnamefont{Chong}} \bibnamefont{and}
  \bibinfo{author}{\bibfnamefont{F.}~\bibnamefont{Sciortino}},
  \bibinfo{journal}{Phy. Rev. E} \textbf{\bibinfo{volume}{69}},
  \bibinfo{pages}{051202} (\bibinfo{year}{2004}).

\bibitem[{\citenamefont{{Chong} and {G{\"o}tze}}(2002)}]{chonggotze}
\bibinfo{author}{\bibfnamefont{S.-H.} \bibnamefont{{Chong}}} \bibnamefont{and}
  \bibinfo{author}{\bibfnamefont{W.}~\bibnamefont{{G{\"o}tze}}},
  \bibinfo{journal}{Phys. Rev. E} \textbf{\bibinfo{volume}{65}},
  \bibinfo{pages}{041503} (\bibinfo{year}{2002}).

\bibitem[{\citenamefont{{Yatsenko} and {Schweizer}}(2007)}]{yatsenko}
\bibinfo{author}{\bibfnamefont{G.}~\bibnamefont{{Yatsenko}}} \bibnamefont{and}
  \bibinfo{author}{\bibfnamefont{K.~S.} \bibnamefont{{Schweizer}}},
  \bibinfo{journal}{J. Chem. Phys.} \textbf{\bibinfo{volume}{126}},
  \bibinfo{pages}{4505} (\bibinfo{year}{2007}).

\bibitem[{\citenamefont{{Fabbian} et~al.}(2000)\citenamefont{{Fabbian}, {Latz},
  {Schilling}, {Sciortino}, {Tartaglia}, and {Theis}}}]{Fab00a}
\bibinfo{author}{\bibfnamefont{L.}~\bibnamefont{{Fabbian}}},
  \bibinfo{author}{\bibfnamefont{A.}~\bibnamefont{{Latz}}},
  \bibinfo{author}{\bibfnamefont{R.}~\bibnamefont{{Schilling}}},
  \bibinfo{author}{\bibfnamefont{F.}~\bibnamefont{{Sciortino}}},
  \bibinfo{author}{\bibfnamefont{P.}~\bibnamefont{{Tartaglia}}},
  \bibnamefont{and} \bibinfo{author}{\bibfnamefont{C.}~\bibnamefont{{Theis}}},
  \bibinfo{journal}{Phys. Rev. E} \textbf{\bibinfo{volume}{62}},
  \bibinfo{pages}{2388} (\bibinfo{year}{2000}).

\bibitem[{\citenamefont{Sciortino and Kob}(2001)}]{Sci01aPRL}
\bibinfo{author}{\bibfnamefont{F.}~\bibnamefont{Sciortino}} \bibnamefont{and}
  \bibinfo{author}{\bibfnamefont{W.}~\bibnamefont{Kob}},
  \bibinfo{journal}{Phys. Rev. Lett.} \textbf{\bibinfo{volume}{86}},
  \bibinfo{pages}{648} (\bibinfo{year}{2001}).

\bibitem[{\citenamefont{{Coluzzi} and {Verrocchio}}(2002)}]{Verrocchio}
\bibinfo{author}{\bibfnamefont{B.}~\bibnamefont{{Coluzzi}}} \bibnamefont{and}
  \bibinfo{author}{\bibfnamefont{P.}~\bibnamefont{{Verrocchio}}},
  \bibinfo{journal}{J. Chem. Phys.} \textbf{\bibinfo{volume}{116}},
  \bibinfo{pages}{3789} (\bibinfo{year}{2002}).

\bibitem[{\citenamefont{{Kob} et~al.}(2002)\citenamefont{{Kob}, {Nauroth}, and
  {Sciortino}}}]{nauroth}
\bibinfo{author}{\bibfnamefont{W.}~\bibnamefont{{Kob}}},
  \bibinfo{author}{\bibfnamefont{M.}~\bibnamefont{{Nauroth}}},
  \bibnamefont{and}
  \bibinfo{author}{\bibfnamefont{F.}~\bibnamefont{{Sciortino}}},
  \bibinfo{journal}{J. Non-Cryst. Sol.} \textbf{\bibinfo{volume}{307}},
  \bibinfo{pages}{181} (\bibinfo{year}{2002}).

\bibitem[{\citenamefont{{van Beest} et~al.}(1990)\citenamefont{{van Beest},
  {Kramer}, and {van Santen}}}]{bks}
\bibinfo{author}{\bibfnamefont{B.~W.~H.} \bibnamefont{{van Beest}}},
  \bibinfo{author}{\bibfnamefont{G.~J.} \bibnamefont{{Kramer}}},
  \bibnamefont{and} \bibinfo{author}{\bibfnamefont{R.~A.} \bibnamefont{{van
  Santen}}}, \bibinfo{journal}{Phys. Rev. Lett.} \textbf{\bibinfo{volume}{64}},
  \bibinfo{pages}{1955} (\bibinfo{year}{1990}).

\bibitem[{\citenamefont{{Zaccarelli} et~al.}(2007)\citenamefont{{Zaccarelli},
  {Sciortino}, and {Tartaglia}}}]{prossimo}
\bibinfo{author}{\bibfnamefont{E.}~\bibnamefont{{Zaccarelli}}},
  \bibinfo{author}{\bibfnamefont{F.}~\bibnamefont{{Sciortino}}},
  \bibnamefont{and}
  \bibinfo{author}{\bibfnamefont{P.}~\bibnamefont{{Tartaglia}}},
  \bibinfo{journal}{in preparation}  (\bibinfo{year}{2007}).

\bibitem[{\citenamefont{{Largo}
  et~al.}(2007{\natexlab{a}})\citenamefont{{Largo}, {Tartaglia}, and
  {Sciortino}}}]{LargoPRE}
\bibinfo{author}{\bibfnamefont{J.}~\bibnamefont{{Largo}}},
  \bibinfo{author}{\bibfnamefont{P.}~\bibnamefont{{Tartaglia}}},
  \bibnamefont{and}
  \bibinfo{author}{\bibfnamefont{F.}~\bibnamefont{{Sciortino}}},
  \bibinfo{journal}{Phys. Rev. E} \textbf{\bibinfo{volume}{76}},
  \bibinfo{pages}{011402} (\bibinfo{year}{2007}{\natexlab{a}}).

\bibitem[{\citenamefont{Kern and D.Frenkel}(2003)}]{Kern_03}
\bibinfo{author}{\bibfnamefont{N.}~\bibnamefont{Kern}} \bibnamefont{and}
  \bibinfo{author}{\bibnamefont{D.Frenkel}}, \bibinfo{journal}{J. Chem. Phys.}
  \textbf{\bibinfo{volume}{{\bf 118}}}, \bibinfo{pages}{9882}
  (\bibinfo{year}{2003}).

\bibitem[{\citenamefont{{Horbach} and {Kob}}(1999)}]{Hor99aPRB}
\bibinfo{author}{\bibfnamefont{J.}~\bibnamefont{{Horbach}}} \bibnamefont{and}
  \bibinfo{author}{\bibfnamefont{W.}~\bibnamefont{{Kob}}},
  \bibinfo{journal}{Phys. Rev. B} \textbf{\bibinfo{volume}{60}},
  \bibinfo{pages}{3169} (\bibinfo{year}{1999}).

\bibitem[{\citenamefont{{Giovambattista}
  et~al.}(2005)\citenamefont{{Giovambattista}, {Stanley}, and
  {Sciortino}}}]{giovambattista}
\bibinfo{author}{\bibfnamefont{N.}~\bibnamefont{{Giovambattista}}},
  \bibinfo{author}{\bibfnamefont{H.~E.} \bibnamefont{{Stanley}}},
  \bibnamefont{and}
  \bibinfo{author}{\bibfnamefont{F.}~\bibnamefont{{Sciortino}}},
  \bibinfo{journal}{Phys. Rev. E} \textbf{\bibinfo{volume}{72}},
  \bibinfo{pages}{031510} (\bibinfo{year}{2005}).

\bibitem[{\citenamefont{{Loerting} and
  {Giovambattista}}(2006)}]{loertingnicolas}
\bibinfo{author}{\bibfnamefont{T.}~\bibnamefont{{Loerting}}} \bibnamefont{and}
  \bibinfo{author}{\bibfnamefont{N.}~\bibnamefont{{Giovambattista}}},
  \bibinfo{journal}{J. Phys.: Condens. Matter} \textbf{\bibinfo{volume}{18}},
  \bibinfo{pages}{919} (\bibinfo{year}{2006}).

\bibitem[{\citenamefont{{Starr} et~al.}(1999)\citenamefont{{Starr},
  {Bellissent-Funel}, and {Stanley}}}]{starrsq}
\bibinfo{author}{\bibfnamefont{F.~W.} \bibnamefont{{Starr}}},
  \bibinfo{author}{\bibfnamefont{M.-C.} \bibnamefont{{Bellissent-Funel}}},
  \bibnamefont{and} \bibinfo{author}{\bibfnamefont{H.~E.}
  \bibnamefont{{Stanley}}}, \bibinfo{journal}{Phys. Rev. E}
  \textbf{\bibinfo{volume}{60}}, \bibinfo{pages}{1084} (\bibinfo{year}{1999}).

\bibitem[{\citenamefont{Hansen and MacDonald}(2006)}]{hansen06}
\bibinfo{author}{\bibfnamefont{J.~P.} \bibnamefont{Hansen}} \bibnamefont{and}
  \bibinfo{author}{\bibfnamefont{I.~R.} \bibnamefont{MacDonald}},
  \emph{\bibinfo{title}{Theory of Simple Liquids}}
  (\bibinfo{publisher}{Academic}, \bibinfo{address}{London},
  \bibinfo{year}{2006}), \bibinfo{edition}{3rd} ed.

\bibitem[{\citenamefont{{Duda} et~al.}(1998)\citenamefont{{Duda}, {Segura},
  {Vakarin}, {Holovko}, and {Chapman}}}]{duda1}
\bibinfo{author}{\bibfnamefont{Y.}~\bibnamefont{{Duda}}},
  \bibinfo{author}{\bibfnamefont{C.~J.} \bibnamefont{{Segura}}},
  \bibinfo{author}{\bibfnamefont{E.}~\bibnamefont{{Vakarin}}},
  \bibinfo{author}{\bibfnamefont{M.~F.} \bibnamefont{{Holovko}}},
  \bibnamefont{and} \bibinfo{author}{\bibfnamefont{W.~G.}
  \bibnamefont{{Chapman}}}, \bibinfo{journal}{J. Chem. Phys.}
  \textbf{\bibinfo{volume}{108}}, \bibinfo{pages}{9168} (\bibinfo{year}{1998}).

\bibitem[{\citenamefont{{Duda}}(1998)}]{duda2}
\bibinfo{author}{\bibfnamefont{Y.}~\bibnamefont{{Duda}}}, \bibinfo{journal}{J.
  Chem. Phys.} \textbf{\bibinfo{volume}{109}}, \bibinfo{pages}{9015}
  (\bibinfo{year}{1998}).

\bibitem[{\citenamefont{Moreno et~al.}(2005)\citenamefont{Moreno, Buldyrev, {La
  Nave}, Saika-Voivod, Sciortino, Tartaglia, and Zaccarelli}}]{Moreno_05}
\bibinfo{author}{\bibfnamefont{A.~J.} \bibnamefont{Moreno}},
  \bibinfo{author}{\bibfnamefont{S.~V.} \bibnamefont{Buldyrev}},
  \bibinfo{author}{\bibfnamefont{E.}~\bibnamefont{{La Nave}}},
  \bibinfo{author}{\bibfnamefont{I.}~\bibnamefont{Saika-Voivod}},
  \bibinfo{author}{\bibfnamefont{F.}~\bibnamefont{Sciortino}},
  \bibinfo{author}{\bibfnamefont{P.}~\bibnamefont{Tartaglia}},
  \bibnamefont{and}
  \bibinfo{author}{\bibfnamefont{E.}~\bibnamefont{Zaccarelli}},
  \bibinfo{journal}{Phys. Rev. Lett.} \textbf{\bibinfo{volume}{{\bf 95}}},
  \bibinfo{pages}{157802} (\bibinfo{year}{2005}).

\bibitem[{\citenamefont{Moreno et~al.}(2006)\citenamefont{Moreno, Saika-Voivod,
  Zaccarelli, Nave, Buldyrev, Tartaglia, and Sciortino}}]{MorenoJCP}
\bibinfo{author}{\bibfnamefont{A.~J.} \bibnamefont{Moreno}},
  \bibinfo{author}{\bibfnamefont{I.}~\bibnamefont{Saika-Voivod}},
  \bibinfo{author}{\bibfnamefont{E.}~\bibnamefont{Zaccarelli}},
  \bibinfo{author}{\bibfnamefont{E.~L.} \bibnamefont{Nave}},
  \bibinfo{author}{\bibfnamefont{S.~V.} \bibnamefont{Buldyrev}},
  \bibinfo{author}{\bibfnamefont{P.}~\bibnamefont{Tartaglia}},
  \bibnamefont{and}
  \bibinfo{author}{\bibfnamefont{F.}~\bibnamefont{Sciortino}},
  \bibinfo{journal}{J. Chem. Phys.} \textbf{\bibinfo{volume}{{\bf 124}}},
  \bibinfo{pages}{204509} (\bibinfo{year}{2006}).

\bibitem[{\citenamefont{Wertheim}(1984)}]{Werth1}
\bibinfo{author}{\bibfnamefont{M.}~\bibnamefont{Wertheim}},
  \bibinfo{journal}{J. Stat. Phys.} \textbf{\bibinfo{volume}{{35}}},
  \bibinfo{pages}{19} (\bibinfo{year}{1984}), \bibinfo{note}{; 1984 {\it J.
  Stat. Phys.} {\bf 35} 35; 1986 {\it J. Chem. Phys.} {\bf 85} 2929}.

\bibitem[{\citenamefont{Sciortino et~al.}(2007)\citenamefont{Sciortino,
  Bianchi, Douglas, and Tartaglia}}]{Sci07a}
\bibinfo{author}{\bibfnamefont{F.}~\bibnamefont{Sciortino}},
  \bibinfo{author}{\bibfnamefont{E.}~\bibnamefont{Bianchi}},
  \bibinfo{author}{\bibfnamefont{J.}~\bibnamefont{Douglas}}, \bibnamefont{and}
  \bibinfo{author}{\bibfnamefont{P.}~\bibnamefont{Tartaglia}},
  \bibinfo{journal}{J. Chem. Phys.} \textbf{\bibinfo{volume}{126}},
  \bibinfo{pages}{194903} (\bibinfo{year}{2007}).

\bibitem[{\citenamefont{Bianchi et~al.}(2007)\citenamefont{Bianchi, {La Nave},
  Tartaglia, and Sciortino}}]{Bia07a}
\bibinfo{author}{\bibfnamefont{E.}~\bibnamefont{Bianchi}},
  \bibinfo{author}{\bibfnamefont{E.}~\bibnamefont{{La Nave}}},
  \bibinfo{author}{\bibfnamefont{P.}~\bibnamefont{Tartaglia}},
  \bibnamefont{and} \bibinfo{author}{\bibfnamefont{F.}~\bibnamefont{Sciortino}}
  (\bibinfo{year}{2007}).

\bibitem[{\citenamefont{{Rino} et~al.}(1993)\citenamefont{{Rino}, {Ebbsj{\"o}},
  {Kalia}, {Nakano}, and {Vashishta}}}]{Rino93}
\bibinfo{author}{\bibfnamefont{J.~P.} \bibnamefont{{Rino}}},
  \bibinfo{author}{\bibfnamefont{I.}~\bibnamefont{{Ebbsj{\"o}}}},
  \bibinfo{author}{\bibfnamefont{R.~K.} \bibnamefont{{Kalia}}},
  \bibinfo{author}{\bibfnamefont{A.}~\bibnamefont{{Nakano}}}, \bibnamefont{and}
  \bibinfo{author}{\bibfnamefont{P.}~\bibnamefont{{Vashishta}}},
  \bibinfo{journal}{Phys. Rev. B} \textbf{\bibinfo{volume}{47}},
  \bibinfo{pages}{3053} (\bibinfo{year}{1993}).

\bibitem[{\citenamefont{{Mousseau} and {Lewis}}(1997)}]{Mou97a}
\bibinfo{author}{\bibfnamefont{N.}~\bibnamefont{{Mousseau}}} \bibnamefont{and}
  \bibinfo{author}{\bibfnamefont{L.~J.} \bibnamefont{{Lewis}}},
  \bibinfo{journal}{Phys. Rev. Lett.} \textbf{\bibinfo{volume}{78}},
  \bibinfo{pages}{1484} (\bibinfo{year}{1997}).

\bibitem[{\citenamefont{{Vollmayr} et~al.}(1996)\citenamefont{{Vollmayr},
  {Kob}, and {Binder}}}]{VolPRB}
\bibinfo{author}{\bibfnamefont{K.}~\bibnamefont{{Vollmayr}}},
  \bibinfo{author}{\bibfnamefont{W.}~\bibnamefont{{Kob}}}, \bibnamefont{and}
  \bibinfo{author}{\bibfnamefont{K.}~\bibnamefont{{Binder}}},
  \bibinfo{journal}{Phys. Rev. B} \textbf{\bibinfo{volume}{54}},
  \bibinfo{pages}{15808} (\bibinfo{year}{1996}).

\bibitem[{\citenamefont{Speedy and Debenedetti}(1994)}]{Spe94}
\bibinfo{author}{\bibfnamefont{R.~J.} \bibnamefont{Speedy}} \bibnamefont{and}
  \bibinfo{author}{\bibfnamefont{P.~G.} \bibnamefont{Debenedetti}},
  \bibinfo{journal}{Mol. Phys.} \textbf{\bibinfo{volume}{81}},
  \bibinfo{pages}{237} (\bibinfo{year}{1994}).

\bibitem[{\citenamefont{Speedy and Debenedetti}(1996)}]{Spe96}
\bibinfo{author}{\bibfnamefont{R.~J.} \bibnamefont{Speedy}} \bibnamefont{and}
  \bibinfo{author}{\bibfnamefont{P.~G.} \bibnamefont{Debenedetti}},
  \bibinfo{journal}{Mol. Phys.} \textbf{\bibinfo{volume}{88}},
  \bibinfo{pages}{1293} (\bibinfo{year}{1996}).

\bibitem[{\citenamefont{Kolafa and Nezbeda}(1987)}]{Kol_87}
\bibinfo{author}{\bibfnamefont{J.}~\bibnamefont{Kolafa}} \bibnamefont{and}
  \bibinfo{author}{\bibfnamefont{I.}~\bibnamefont{Nezbeda}},
  \bibinfo{journal}{Mol. Phys.} \textbf{\bibinfo{volume}{{\bf 61}}},
  \bibinfo{pages}{161} (\bibinfo{year}{1987}).

\bibitem[{\citenamefont{Ford et~al.}(2004)\citenamefont{Ford, Auerbach, and
  Monson}}]{Ford_04}
\bibinfo{author}{\bibfnamefont{M.~H.} \bibnamefont{Ford}},
  \bibinfo{author}{\bibfnamefont{S.~M.} \bibnamefont{Auerbach}},
  \bibnamefont{and} \bibinfo{author}{\bibfnamefont{P.~A.}
  \bibnamefont{Monson}}, \bibinfo{journal}{J. Chem. Phys.}
  \textbf{\bibinfo{volume}{{\bf 121}}}, \bibinfo{pages}{8415}
  (\bibinfo{year}{2004}).

\bibitem[{\citenamefont{{Romano} et~al.}(2007)\citenamefont{{Romano},
  {Tartaglia}, and {Sciortino}}}]{romanoJPCM}
\bibinfo{author}{\bibfnamefont{F.}~\bibnamefont{{Romano}}},
  \bibinfo{author}{\bibfnamefont{P.}~\bibnamefont{{Tartaglia}}},
  \bibnamefont{and}
  \bibinfo{author}{\bibfnamefont{F.}~\bibnamefont{{Sciortino}}},
  \bibinfo{journal}{J. Phys.: Condens. Matter} \textbf{\bibinfo{volume}{19}},
  \bibinfo{pages}{F2101+} (\bibinfo{year}{2007}).

\bibitem[{\citenamefont{Vega and Monson}(1998)}]{Monson_98}
\bibinfo{author}{\bibfnamefont{C.}~\bibnamefont{Vega}} \bibnamefont{and}
  \bibinfo{author}{\bibfnamefont{P.~A.} \bibnamefont{Monson}},
  \bibinfo{journal}{J. Chem. Phys.} \textbf{\bibinfo{volume}{{\bf 109}}},
  \bibinfo{pages}{9938} (\bibinfo{year}{1998}).

\bibitem[{\citenamefont{Stauffer and Aharony}(1992)}]{Sta92book}
\bibinfo{author}{\bibfnamefont{D.}~\bibnamefont{Stauffer}} \bibnamefont{and}
  \bibinfo{author}{\bibfnamefont{A.}~\bibnamefont{Aharony}},
  \emph{\bibinfo{title}{Introduction to Percolation Theory}}
  (\bibinfo{publisher}{Taylor and Francis}, \bibinfo{address}{London},
  \bibinfo{year}{1992}), \bibinfo{edition}{2nd} ed.

\bibitem[{\citenamefont{Coniglio and Klein}(1980)}]{coniglio-klein}
\bibinfo{author}{\bibfnamefont{A.}~\bibnamefont{Coniglio}} \bibnamefont{and}
  \bibinfo{author}{\bibfnamefont{W.}~\bibnamefont{Klein}}, \bibinfo{journal}{J.
  Phys. A} \textbf{\bibinfo{volume}{13}}, \bibinfo{pages}{2775}
  (\bibinfo{year}{1980}).

\bibitem[{\citenamefont{{Xu} et~al.}(2005)\citenamefont{{Xu}, {Kumar},
  {Buldyrev}, {Chen}, {Poole}, {Sciortino}, and {Stanley}}}]{ScioPNAS}
\bibinfo{author}{\bibfnamefont{L.}~\bibnamefont{{Xu}}},
  \bibinfo{author}{\bibfnamefont{P.}~\bibnamefont{{Kumar}}},
  \bibinfo{author}{\bibfnamefont{S.~V.} \bibnamefont{{Buldyrev}}},
  \bibinfo{author}{\bibfnamefont{S.-H.} \bibnamefont{{Chen}}},
  \bibinfo{author}{\bibfnamefont{P.~H.} \bibnamefont{{Poole}}},
  \bibinfo{author}{\bibfnamefont{F.}~\bibnamefont{{Sciortino}}},
  \bibnamefont{and} \bibinfo{author}{\bibfnamefont{H.~E.}
  \bibnamefont{{Stanley}}}, \bibinfo{journal}{Proc. Nat. Ac. Sci.}
  \textbf{\bibinfo{volume}{102}}, \bibinfo{pages}{16558}
  (\bibinfo{year}{2005}).

\bibitem[{\citenamefont{{Kumar} et~al.}(2007)\citenamefont{{Kumar}, {Buldyrev},
  {Becker}, {Poole}, {Starr}, and {Stanley}}}]{pradeepPNAS}
\bibinfo{author}{\bibfnamefont{P.}~\bibnamefont{{Kumar}}},
  \bibinfo{author}{\bibfnamefont{S.~V.} \bibnamefont{{Buldyrev}}},
  \bibinfo{author}{\bibfnamefont{S.~R.} \bibnamefont{{Becker}}},
  \bibinfo{author}{\bibfnamefont{P.~H.} \bibnamefont{{Poole}}},
  \bibinfo{author}{\bibfnamefont{F.}~\bibnamefont{{Starr}}}, \bibnamefont{and}
  \bibinfo{author}{\bibfnamefont{H.~E.} \bibnamefont{{Stanley}}},
  \bibinfo{journal}{Proc. Nat. Ac. Sci.} \textbf{\bibinfo{volume}{104}},
  \bibinfo{pages}{9575} (\bibinfo{year}{2007}).

\bibitem[{\citenamefont{{Poole} et~al.}(2005)\citenamefont{{Poole},
  {Saika-Voivod}, and {Sciortino}}}]{st2jpcm}
\bibinfo{author}{\bibfnamefont{P.~H.} \bibnamefont{{Poole}}},
  \bibinfo{author}{\bibfnamefont{I.}~\bibnamefont{{Saika-Voivod}}},
  \bibnamefont{and}
  \bibinfo{author}{\bibfnamefont{F.}~\bibnamefont{{Sciortino}}},
  \bibinfo{journal}{J. Phys.: Condens. Matter} \textbf{\bibinfo{volume}{17}},
  \bibinfo{pages}{L431} (\bibinfo{year}{2005}).

\bibitem[{\citenamefont{{Poole} et~al.}(1992)\citenamefont{{Poole},
  {Sciortino}, {Essmann}, and {Stanley}}}]{Poo92}
\bibinfo{author}{\bibfnamefont{P.~H.} \bibnamefont{{Poole}}},
  \bibinfo{author}{\bibfnamefont{F.}~\bibnamefont{{Sciortino}}},
  \bibinfo{author}{\bibfnamefont{U.}~\bibnamefont{{Essmann}}},
  \bibnamefont{and} \bibinfo{author}{\bibfnamefont{H.~E.}
  \bibnamefont{{Stanley}}}, \bibinfo{journal}{Nature}
  \textbf{\bibinfo{volume}{360}}, \bibinfo{pages}{324} (\bibinfo{year}{1992}).

\bibitem[{\citenamefont{{Largo}
  et~al.}(2007{\natexlab{b}})\citenamefont{{Largo}, {Starr}, and
  {Sciortino}}}]{Largo_07}
\bibinfo{author}{\bibfnamefont{J.}~\bibnamefont{{Largo}}},
  \bibinfo{author}{\bibfnamefont{F.~W.} \bibnamefont{{Starr}}},
  \bibnamefont{and}
  \bibinfo{author}{\bibfnamefont{F.}~\bibnamefont{{Sciortino}}},
  \bibinfo{journal}{Langmuir} \textbf{\bibinfo{volume}{23}},
  \bibinfo{pages}{5896} (\bibinfo{year}{2007}{\natexlab{b}}).

\end{thebibliography}

\end{document}